\documentclass[aps, reprint, superscriptaddress, longbibliography]{revtex4-1}
\usepackage{soul}
\usepackage{lineno}
\usepackage{graphicx}
\usepackage{dcolumn}   
\usepackage{bm}        
\usepackage{amssymb}   
\usepackage{dsfont}
\usepackage{relsize}
\usepackage{amsmath,stmaryrd}
\usepackage{blkarray, multirow, graphicx, diagbox, color, colortbl}
\usepackage{bbm} 
\usepackage{glossaries}
\usepackage{hyphenat}
\usepackage{ifthen}
\usepackage[colorlinks, linkcolor = blue, citecolor = blue, filecolor = black, urlcolor = blue]{hyperref}
\usepackage{xkeyval}
\usepackage{moreverb}
\usepackage{rotating}
\usepackage{wrapfig}
\usepackage{xspace}
\usepackage{nicefrac}
\usepackage[]{units}
\usepackage{physics}
\usepackage{booktabs}
\usepackage{braket}
\usepackage[inline]{enumitem}
\usepackage{tabto}
\usepackage{listings}
\usepackage{xstring}
\usepackage{url}
\usepackage[section]{placeins}
\usepackage[normalem]{ulem}
\usepackage{mathrsfs}
\usepackage{mathtools}
\usepackage[scr=boondoxo]{mathalpha} 

\def\ReplaceStr#1{%
	\IfSubStr{#1}{p}{%
		\StrSubstitute{#1}{p}{.}}{#1}}

\usepackage[capitalise]{cleveref} %
\usepackage{chemformula}
\usepackage{algorithm}

\DeclareMathOperator*{\argmax}{arg\,max}

\usepackage{tikz}

\usetikzlibrary{calc}

\usepackage[dvipsnames]{xcolor}
\definecolor{mygreen}{HTML}{009000}
\definecolor{myorange}{HTML}{ff6600}
\definecolor{mypurple}{HTML}{6F00FF}
\definecolor{electriclime}{rgb}{0.8, 1.0, 0.0}

\definecolor{block}{RGB}{0,162,232}

\usepackage{tikz}

\definecolor{cibord}{HTML}{ff7f31}   
\definecolor{cifill}{HTML}{fff7f3}      

\newcommand{\orangecircle}{%
  \tikz[baseline=-.75ex] \draw[cibord, fill=cifill, line width=.5pt] (.1,.015) circle (1.ex);%
}

\usetikzlibrary{shapes.geometric}

\definecolor{bluebord}{RGB}{63,79,162}  
\definecolor{bluefill}{RGB}{214,215,236} 

\newcommand{\bluestar}{%
  \tikz[baseline=-.75ex]%
    \node[
      star,
      star points=5,
      star point ratio=1.6,   
      draw=bluebord,
      fill=bluefill,
      line width=.5pt,       
      inner sep=0pt,
      minimum size=2.5ex      
    ] {};
}
\definecolor{sqbord}{HTML}{8d6bb8}
\definecolor{sqfill}{HTML}{f4f2f4}

\newcommand{\purplesquare}{
  \tikz[baseline=-.75ex]
    \node[
      draw=sqbord,
      fill=sqfill,
      line width=.5pt,
      minimum size=2.ex,   
      inner sep=0pt         
    ] {};
}

\definecolor{tribord}{HTML}{529d3f}
\definecolor{trifill}{HTML}{f5f7f4}

\newcommand{\greentriangle}{%
  \tikz[baseline=-.5ex] \draw[
      draw=tribord,
      fill=trifill,
      line width=.5pt
    ]
    (0, 1.2ex) -- (-1ex, -0.6ex) -- (1ex, -0.6ex) -- cycle;%
}

\def\blockaux#1(#2,#3)#4(#5,#6){%
  \draw[fill={#1}, draw=none]
  let \p1=(#2,#3),
      \p2=(#5,#6),
      \p3=(#2+#5,#3+#6),
      \p4=(#2+#5/2,#3+#6/2)
  in
    (\p1) rectangle (\p3)
    (\p4) node {\centering $#4$}
  ;%
}

\newacronym{OBC}{OBC}{open boundary condition}
\newacronym{CPTP}{CPTP}{completely positive and trace-preserving}
\newacronym{TFIM}{TFIM}{transverse-field Ising model}
\newacronym{QME}{QME}{quantum Mpemba effect}
\newacronym{ETH}{ETH}{eigenstate thermalization hypothesis}
\newacronym[
  shortplural={QRTs},
  longplural={Quantum Resource Theories}
]{QRT}{QRT}{Quantum Resource Theory}
\newacronym[
  shortplural={RTs},
  longplural={Resource Theories}
]{RT}{RT}{Resource Theory}
\newacronym{KL}{KL}{Kullback-Leibler}
\newacronym{TO}{TO}{Thermal Operation}
\newacronym{QFI}{QFI}{Quantum Fisher Information}
\newacronym{LOCC}{LOCC}{local operations and classical communication}
\newacronym{PBC}{PBC}{periodic boundary condition}
\newacronym{CG}{CG}{Clebsch-Gordan}

\graphicspath{{figures/}}
\def\affilA{School of Physics, Trinity College Dublin, Dublin 2, Ireland} 
\def\affilB{Trinity Quantum Alliance, Unit 16, Trinity Technology and Enterprise Centre, Pearse Street, D02 YN67, Dublin 2, Ireland}  
 
\def\affilD{Institut für Theoretische Physik, Universität zu Köln, Zülpicher Strasse 77, 50937 Köln, Germany} 
\def\affilE{Duke Quantum Center and Department of Physics, Duke University, Durham, NC 27708, USA} 
\def\affilF{Department of Electrical and Computer Engineering, Duke University, Durham, NC 27708, USA} 
\begin{document}
\def\thetitle{Resource{-}Theoretical Unification of Mpemba Effects: Classical and Quantum} 
\author{Alessandro Summer}
\email{summera@tcd.ie}
\affiliation{\affilA}
\affiliation{\affilB}

\author{Mattia Moroder}
\email{moroderm@tcd.ie}
\affiliation{\affilA}
\affiliation{\affilB}

\author{Laetitia P. Bettmann}
\email{bettmanl@tcd.ie}
\affiliation{\affilA}

\author{Xhek Turkeshi}
\email{turkeshi@thp.uni-koeln.de}
\affiliation{\affilD}

\author{Iman Marvian}
\email{iman.marvian@duke.edu}
\affiliation{\affilE}
\affiliation{\affilF}

\author{John Goold}
\email{gooldj@tcd.ie}
\affiliation{\affilA}
\affiliation{\affilB}
%
\title{\thetitle}
\begin{abstract}
The Mpemba effect originally referred to the observation that, under certain thermalizing dynamics, initially hotter samples can cool faster than colder ones. This effect has since been generalized to other anomalous relaxation behaviors even beyond classical domains, such as symmetry restoration in quantum systems.
This work demonstrates that 
resource theories, widely employed in information theory, provide a unified organizing principle to frame Mpemba physics. 
We show how the conventional thermal Mpemba effect arises naturally from the resource theory of athermality, while its symmetry-restoring counterpart is fully captured by the resource theories of asymmetry.
Leveraging the framework of modes of asymmetry, we demonstrate that the Mpemba effect due to symmetry restoration is governed by the initial overlap with the slowest symmetry-restoring mode -- mirroring the role of the slowest Liouvillian eigenmode in thermal Mpemba dynamics.
Through this resource-theoretical formalism, we uncover the connection between these seemingly disparate effects and show that the dynamics of thermalization naturally splits into a symmetry-respecting and a symmetry-breaking term. 
%
%
%
%
%
%
\end{abstract}
\maketitle
\section{Introduction}
\label{sec:introduction}
The Mpemba effect originally described the anomalous situation of a hot system cooling down faster than a warm one.
First investigated by Mpemba and Osborne in water in 1969~\cite{Mpemba1969}, it has since then been observed in various physical systems ranging from clathrate hydrates~\cite{Ahn2016} to crystallization processes in polymers~\cite{Hu2018} and magnetic transitions in alloys~\cite{Chaddah2010}. 
Following its precise mathematical characterization in classical systems coupled to Markovian environments~\cite{Lu2017}, the effect was further investigated theoretically~\cite{Klich2019, Gal2020, Teza2023, Teza2025} and experimentally~\cite{Kumar2020, Kumar2022, Malhotra2024} for colloidal systems. 
{Two further important directions in the study of the classical Mpemba effect concern driven granular gases~\cite{Lasanta2017, Biswas2020, Teza2025} and phase transitions~\cite{Yang2020,Zhang2022}. The latter, despite pertaining to the phenomenon as originally observed in water, remains the most challenging and least understood case~\cite{Jeng2006,buridge2020observing}.}

In recent years, the theoretical framework was extended to quantum systems~\cite{Nava2019, Carollo2021, Bao2022, Kochsiek2022, Ivander2023, Wang2024, Aharony2024, Moroder2024, Strachan2024, Medina2024, Xu2025, Westhoff2025, Beato2026}.
In this context, a related phenomenon has been observed in the dynamical restoration of symmetries: states that initially display stronger symmetry breaking can \emph{locally}~\footnote{
Throughout this paper we consider dynamics on a composite Hilbert space $\mathscr H = \mathscr H_s\otimes\mathscr H_e$, where $\mathscr H_s$ (system) and $\mathscr H_e$ (environment) are always distinguished. In the open-dynamics case 
$\mathscr H_e$ describes an external bath; in the {closed}-dynamics case $\mathscr H_e$ is 
the complementary subsystem within an overall isolated $\mathscr H$. \emph{Local} symmetry restoration then means that, after evolving under the joint dynamics, we apply the partial trace 
$\Tr_e$ and study symmetry recovery on the reduced state $\hat \rho_s = \Tr_e[\hat \sigma]$.}
restore it faster than states with initially weaker symmetry breaking~\cite{Ares2023Nat, Ares2025}.
We refer to this phenomenon as the \emph{symmetry} Mpemba effect, to differentiate it from the conventional \emph{thermal} Mpemba effect pertaining to thermalization dynamics.
\begin{figure}
    \centering
    \includegraphics[width=\linewidth]{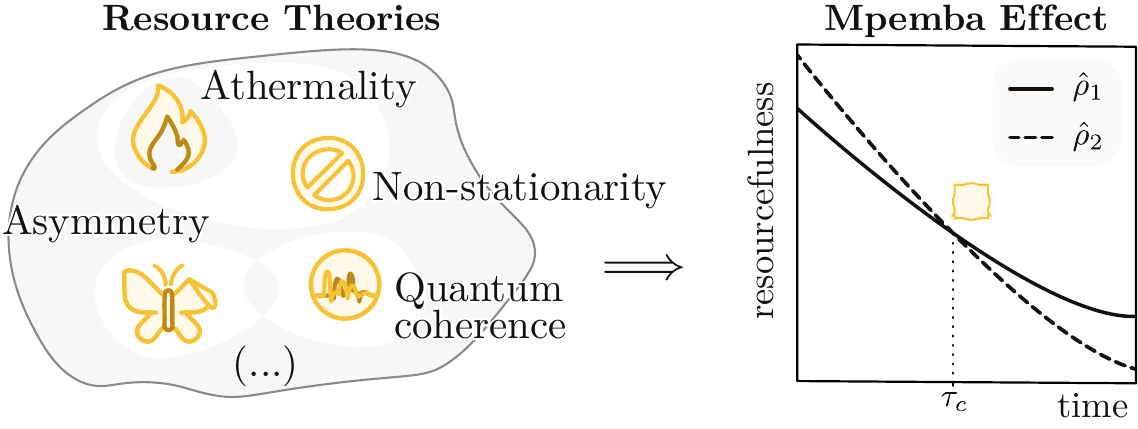}
    \caption{In a resource-theoretic framework, the Mpemba effect occurs when a state that initially possesses more of a given resource depletes that resource faster than a less resourceful state, under the evolution by the same free operation, so that their resource monotones cross. This single picture unifies a variety of anomalous equilibration phenomena (for example, restoring thermal equilibrium or symmetry in classical and quantum systems). 
    Mpemba physics then becomes the study of why different initial states dissipate resources at different rates, and how we can harness those differences to engineer exotic effects such as ultrafast cooling. In this article, we apply this analysis to the specific resource theories shown in the schematic.
    }
    \vspace{-.6cm}
    \label{fig:intro}
\end{figure}

In this article, we establish a framework to understand both the thermal and symmetry Mpemba effects on the same footing using \glspl{RT}~\cite{Coecke2016, Chitambar2019} which are widely used in quantum information theory. 
A Mpemba effect arises when two resourceful states evolve under the same dynamics, and the more resourceful one dissipates the resource faster than the less resourceful one, causing their resource monotones to cross.
This description allows us to capture the essence of both effects within a single conceptual framework.
Building upon this, we employ the framework of modes of asymmetry~\cite{Marvian2014b} to explain the occurrence of the symmetry Mpemba effect. Analogous to how the thermal Mpemba effect is understood in terms of overlaps with slow relaxation modes~\cite{Lu2017}, we demonstrate that the symmetry Mpemba effect also arises when strongly symmetry-broken initial states have small (or vanishing) overlap with the slowest symmetry-restoring mode.
Within this framework, we study various examples revealing that the thermal Mpemba effect can also occur in subsystems of unitarily evolving systems, while 
the symmetry Mpemba effect can manifest in quantum systems under Lindblad evolution and in purely
classical Markovian dynamics
, as well as circuits with Abelian (continuous) and non-Abelian symmetries. In this context, we emphasize that the specific type of Mpemba effect observed is dictated by the chosen measure for quantifying the relaxation process, rather than being an intrinsic property of the physical system itself.
Finally, in the presence of arbitrary symmetries, we highlight the connection between the non-equilibrium free energy, commonly used for the thermal Mpemba effect, and the relative entropy of asymmetry (also known as entanglement asymmetry), a measure for the asymmetry of a state.\\
To guide the reader through the paper, Mpemba effects originated from different \gls{RT} are indicated via different symbols: \purplesquare\hspace{-0.25em} for athermality, \orangecircle\hspace{0.125em} for asymmetry, \greentriangle\hspace{0.125em} for symmetry-respecting athermality, and \bluestar
for non-stationarity.

This article is organized as follows: First, in \cref{sec:resource_theories}, we concisely present some basic concepts regarding \glspl{RT}.
Subsequently, in \cref{sec:thermal:mpemba} we explore the occurrence of the thermal Mpemba effect in various classical and quantum setups from the perspective of \glspl{RT}.
Then we introduce the modes of asymmetry in \cref{sec:symmetry:mpemba} and employ them to identify fast-equilibrating states in the context of the symmetry Mpemba effect.
In \cref{sec:unified:mpemba} we provide a unified description of the thermal and the symmetry Mpemba effect and finally in \cref{sec:conclusion} we summarize our findings and point towards future applications.

\subsection{Resource Theories}
\label{sec:resource_theories}
%
%
\acrfullpl{RT} (see Ref.~\cite{Chitambar2019} for a review)  provide the natural language for understanding and unifying diverse Mpemba effects: they formalize which state transformations are possible when only certain \emph{free} operations are allowed, and thereby translate relaxation processes into the decay of a resource monotone~\cite{Goold2016}.
In a nutshell, a resource theory is defined by two foundational ingredients: a set of free states and a class of free operations. Free operations are \gls{CPTP} maps that are considered accessible at no cost, either because they are physically implementable with minimal resources or because they reflect fundamental constraints of the theory. Similarly, the set of free states, denoted $\mathscr{F} \subset \mathscr{S}(\mathscr{H}_s)$ (where $\mathscr{S}(\mathscr{H}_s)$ indicates the set of density operators on the Hilbert space $\mathscr{H}_s$), consists of states that can be prepared at no cost. A minimal requirement that any resource theory must satisfy is that the set of free states is closed under free operations (as defined later in this section); that is, for any free state $\hat{\pi} \in \mathscr{F}$ and any free operation $\mathcal{E}$, we have $\mathcal{E}[\hat{\pi}] \in \mathscr{F}$. This closure condition ensures that no resource can be generated from within the free set. 
%
%
All states outside the free set, i.e., $\mathscr{S}(\mathscr{H}_s) \setminus \mathscr{F}$~\footnote{We indicate by $\mathscr{S}(\mathscr{H}_s)$ the set of density matrices on $\mathscr{H}_s$.}, are called resourceful or non-free. These are states that cannot be obtained from any free state using only free operations. Their preparation requires access to non-free operations, and they are typically the carriers of some operational advantage within the theory. 
The presence and quantification of such states lie at the core of resource-theoretic approaches.

To familiarize with these concepts, we recall two prominent examples of quantum resource theories: {the first defined one, bipartite entanglement and classical ones, athermality in classical systems.}
In the theory of bipartite entanglement, the free states are separable states, and the free operations are \gls{LOCC}~\cite{Horodecki2009}. Entangled states lie outside the free set and enable tasks such as quantum teleportation or the violation of Bell inequalities, which are impossible under \gls{LOCC} alone. 
{
A fully classical example is provided by the \gls{RT} of athermality for classical systems.
Here, one considers a system in contact with a thermal environment at inverse temperature $\beta$.
The free operations, also called \glspl{TO}, are stochastic maps with fixed point given by the Gibbs distribution $(\boldsymbol{\pi}_\beta)_i\propto e^{-\beta E_i}$, which is also the only free state~\cite{Faist2015}.
Any state that departs from $\boldsymbol{\pi}_\beta$ therefore possesses athermality, representing a source of non-equilibrium free energy that can be harnessed to perform useful work or drive transformations that are otherwise forbidden under thermal operations.
We will analyze this theory more in depth in \Cref{sec:resource_athermality}, seeing how it naturally extends to quantum systems too.
}


%
A resource monotone is a function $M:\mathscr{S}(\mathscr{H}_s)\to[0,\infty)$ satisfying two key properties:
\begin{enumerate}
\item Vanishing for free states: $\hat\rho\in\mathscr{F}\Rightarrow  M(\hat\rho)=0\,.$

\item Monotonicity under free operations: for every state $\hat\rho$ and every $\mathcal{E}\in\mathscr{O}$,
\begin{equation}
M(\mathcal{E}[\hat\rho])\le M(\hat\rho) \,,
\label{eq:monotone_monotonicity}
\end{equation}
where {the set of} free operations $\mathscr{O}$ consist of all \gls{CPTP} maps $\mathcal{E}:\mathscr{H}_s\to\mathscr{H}_s$ that preserve the free set.
\end{enumerate}

A Mpemba effect manifests when two initial states $\hat\rho_1$ and $\hat\rho_2$ satisfying $M(\hat\rho_1)>M(\hat\rho_2)$ evolve under the same free operation $\mathcal{E}_t$ so that
\begin{equation}
M\bigl(\mathcal{E}_t[\hat\rho_1]\bigr) < M\bigl(\mathcal{E}_t[\hat\rho_2]\bigr)
\quad\text{for some }t>0 \,,
\end{equation}
indicating that the more resourceful state has dissipated the resource faster.

To quantify resourcefulness, we adopt the family of Petz-R\'{e}nyi $\alpha$-divergences~\cite{Mosonyi2011}
\begin{equation}
M_\alpha(\hat\rho)
=\min_{\hat\pi\in\mathscr{F}}
S_\alpha\bigl(\hat\rho \,\|\,\hat\pi\bigr)\,,
\quad \alpha\in(0,1)\cup(1,2]\,,
\label{eq:monotone:quantum:Renyi:divergence}
\end{equation}
where
\begin{equation}
\begin{split}
    S_\alpha(\hat\rho \,\|\, \hat\pi) =&
    \frac{1}{\alpha - 1}\ln \mathrm{Tr}[\hat\rho^\alpha\, \hat\pi^{\,1-\alpha}]\,,
    \label{ eq:quantum:Renyi:divergence} 
\end{split}
\end{equation}
obeys the data-processing inequality under any \gls{CPTP} map~\cite{Mosonyi2011, Chitambar2019} and recovers the quantum relative entropy of resource for $\alpha\rightarrow 1$, i.e.,
\begin{equation}
M(\hat\rho)
= \min_{\hat\pi\in\mathscr{F}}
S\bigl(\hat\rho \,\|\, \hat\pi\bigr) \,,
\label{eq:quantum:relative:entropy} 
\end{equation}
where $S\bigl(\hat\rho \,\|\, \hat\pi\bigr) = \Tr\,\![ \hat\rho\,(\ln \hat\rho -  \ln \hat\pi) ]$ is the quantum relative entropy, i.e., the quantum generalization of the \gls{KL}-divergence.
Physically, $M_\alpha$ measures the divergence of $\hat\rho$ with respect to the closest free state in the R\'{e}nyi sense.
{
We note that \cref{eq:monotone:quantum:Renyi:divergence,eq:quantum:relative:entropy} follow the standard entropic/divergence route to construct monotones in general resource theories (vanish on the free states and nonincreasing under the free operations, see Ref.~\cite{Gour2008}).
}
\section{The thermal Mpemba effect}
\label{sec:thermal:mpemba}
The first rigorous mathematical description of the thermal Mpemba effect was given in 2017 by Lu and Raz, who studied discrete-state continuous-time Markovian processes. 
In~\cite{Lu2017}, the dynamics is generated by a classical Liouvillian $\hat{\mathcal{L}}^\mathrm{cl}$, whose eigenvalues $\lambda_k$ are non-positive, and it is convenient to order them in decreasing order of their real part as $0 = \lambda_1 > \Re(\lambda_2) \geq \Re(\lambda_3) \geq\cdots$.
The classical Liouville equation (see \cref{subs:example:classical:open}) dictates that the contribution to the dynamics from all eigenmodes $\mathbf{r}_k$ of $\hat{\mathcal{L}}^\mathrm{cl}$ decay exponentially in time with a rate given by the real part of the corresponding eigenvalue, except for $\mathbf{r}_1$ corresponding to $\lambda_1=0$ that describes the thermal steady state.
Thus, at long times, the equilibration speed of typical states will be proportional to $\exp(\Re(\lambda_2)t)$.
%
%
Instead, special initial states having zero overlap with the so-called slowest-decaying mode $\mathbf{r}_2$ will equilibrate exponentially faster, with speed $\propto \exp(\Re(\lambda_3)t)$, provided that $\Re(\lambda_3)\neq \Re(\lambda_2)$.
Moreover, if a faster equilibrating state happens to be initially further away from the steady state than a typical state, then the distance of the two states measured with some function $M$ will cross in time and this is called a strong Mpemba effect.
If the fast equilibrating states have small but non-vanishing overlap with the slowest-decaying mode, the speedup will not be exponential, and the crossing indicates a weaker Mpemba effect.
To set the stage for the following examples, we first review athermality, the \gls{RT} relevant to the thermal Mpemba effect.

%
\subsection{Athermality as a Resource}
\label{sec:resource_athermality}
Athermality, {as anticipated for classical systems in \Cref{sec:resource_theories},} a state's deviation from thermal equilibrium $\hat\pi_\beta$, carries useful non-equilibrium free energy
$\Delta F(\hat\rho) = F(\hat\rho)-F(\hat\pi_\beta)$, where $F(\hat\rho)=\Tr\bigl[\hat H_s\,\hat\rho\bigr]
-\beta^{-1}S(\hat\rho)=\beta^{-1}S(\hat\rho\,\|\,\hat\pi_\beta)$~\cite{Donald1987}. 
In the resource theory of athermality, it quantifies the resource that can be converted into work using thermal operations~\cite{Brandao2013, Lostaglio2015} {(defined more precisely below). 
These consist of operations that can be realized by coupling the system to an ancilla prepared in the Gibbs state at inverse temperature $\beta$, applying an energy-preserving unitary, and tracing out the ancilla afterwards, allowing heat exchange while preserving overall energy conservation.}
Framing the thermal Mpemba effect in this \gls{RT} underscores what drives the equilibration process for the thermal Mpemba effect, even though, as with any non-equilibrium process, this free energy ultimately dissipates irreversibly into the bath 
rather than being fully harvested as work~\cite{Esposito2010, Seifert2012}.\\
In this framework, the only free state is the thermal state at the inverse temperature $\beta$ set by the environment (throughout this article, we will consider $k_B=\hbar=1$); any other state is considered resourceful.
%
A quantum channel $\mathcal{E}:\mathscr{B}(\mathscr{H}_s)\rightarrow\mathscr{B}(\mathscr{H}_s)$  (where $\mathscr{B}(\mathscr{H}_s)$ denotes the set of bounded operators on $\mathscr{H}_s$) is a free operation, denoted \gls{TO}, if there exists~\cite{Lostaglio2019}
\begin{itemize}
  \item a thermal bath in a Gibbs state $\hat\pi_\beta = {e^{-\beta \hat{H}_e}}/{Z_e}$ with arbitrary Hamiltonian $\hat{H}_e$ and fixed inverse temperature $\beta$,
  \item a global energy-conserving unitary $\hat U\in \mathscr{B}(\mathscr{H}_s\otimes\mathscr{H}_e)$ such that $[\hat U, \hat H_s \otimes \hat H_e] = 0$,
\end{itemize}
where the action of $\mathcal{E}$ on a system state $\hat\rho$ is given by
\begin{equation}
\mathcal{E}[\hat \rho] = \mathrm{Tr}_e[ \hat U (\hat \rho \otimes \hat \pi_\beta) \hat U^\dagger ].
\end{equation}
%
%
%
%
%
%
Trivially, \cref{eq:monotone:quantum:Renyi:divergence} becomes
\begin{equation}
    \label{eq:monotone:quantum:Renyi:divergence:athermality}
    M_\alpha (\hat \rho(t)) = S_\alpha(\hat\rho_s(t) \,\|\,\hat\pi_\beta) \,,
\end{equation}
which for $\alpha\rightarrow 1$ recovers the  \emph{relative entropy of athermality}
\begin{equation}
     M (\hat \rho(t))= S(\hat\rho_s(t) \,\|\,\hat\pi_\beta) \,.
     \label{eq:quantum-relative-entropy:athermality}
\end{equation}
For states that are diagonal in the energy eigenbasis, (i.e., classical states), this reduces to the \gls{KL}-divergence~\cite{MacKay2002}.
An important property is that a state $\hat{\rho}$ can be transformed into another state $\hat{\sigma}$ via \glspl{TO} only if $\hat{\rho}$ thermomajorizes $\hat{\sigma}$~\cite{Ng2018}.
Notably, it was shown that for classical Markovian thermal operations, thermomajorization can be used to determine whether the Mpemba effect occurs for \emph{all} monotone functions~\cite{Vanvu2025}.
In the next section, we will explore examples of classical and quantum thermal Mpemba effects from the perspective of the resource theory of athermality.
More generally, one can consider open system dynamics generated by a Lindbladian $\mathcal L$ whose unique steady state $\hat\pi$ need not be Gibbsian, as often studied in the context of the Mpemba effect~\cite{Carollo2021,Kochsiek2022, Wang2024, Nava2024}.  
In that setting, the relevant resource theory is non-stationarity, in which a state's divergence with respect to $\hat\pi$, i.e. $S(\hat\rho(t)\,\|\,\hat \pi)$,  quantifies its resourcefulness. We discuss the framework of non-stationarity, {which is directly applicable to a class of Mpemba effects in driven granular gases~\cite{Lasanta2017, Biswas2020, Teza2025},} in \cref{app:Mpemba:Liouvillian:symmetry}. To illustrate the role of athermality in the thermal Mpemba effect, we now turn to two paradigmatic examples: a classical system akin to that in Ref.~\cite{Lu2017}, and a quantum counterpart inspired by Ref.~\cite{Moroder2024}. 
In \cref{app:ETH:thermal:Mpemba}, we instead demonstrate how this phenomenon can arise in isolated settings.
{In Ref.~\cite{Alyuruk2025}, an alternative resource-theoretic interpretation of the thermal Mpemba effect is developed, using thermomajorization to analyze the roles of correlations, coherence, non-Markovianity, and dimensionality in both classical and quantum relaxation dynamics, with water serving as a central example.}
\subsection{Example 1: Thermalization of a Spin Chain}
\label{subs:example:classical:open}
%
%
The quantum resource-theoretic framework for athermality introduced above can also be applied to classical dynamics, such as for instance discrete-state, continuous-time Markov processes~\cite{Lu2017, Teza2023a, Teza2025}.
In this context, the monotone for athermality \cref{eq:monotone:quantum:Renyi:divergence:athermality}, i.e. the \gls{KL}-divergence, has already been studied. The \gls{KL}-divergence serves as a generalized free energy: it is non-increasing under any stochastic map preserving the steady state $\boldsymbol{\pi}_\beta$, and its decay is due to the depletion of classical athermality.

Consider a classical system whose state is specified by a probability distribution 
\begin{equation}
    \mathbf{p}(t) = \begin{pmatrix}
p_1(t) \\
p_2(t) \\
\vdots \\
p_L(t) 
\end{pmatrix} \,,
\label{eq:classical:prob:distr}
\end{equation}
where the subscript $i = 1, 2, \dots L$ labels discrete microstates.
For the classical Markovian Liouvillian dynamics
\begin{equation}
     \frac{\mathrm{d} \mathbf{p}(t)}{\mathrm{d}t} = \hat{\mathcal{L}}^\mathrm{cl}\, \mathbf{p}(t) \,,
    \label{eq:classical:liouville:equation}
\end{equation}
the fixed point is the thermal distribution  $\mathbf{p}(t\to \infty) = \boldsymbol\pi_\beta$ at the inverse temperature $\beta$ if the classical Liouvillian matrix $ \hat{\mathcal{L}}^\mathrm{cl}$ is 
\begin{equation}
\hat{\mathcal{L}}^\mathrm{cl}_{ii'} =
\begin{cases}
W_{ii'}, & i \neq i'\, , \\
-\!\sum_{i \neq i'} W_{ii'}, & i = i'\,.
\end{cases}
\label{eq:classical:liouvillian}
\end{equation}
Here, $W_{ii'}$ satisfies the classical detailed balance condition $W_{ii'}/W_{i'i}=e^{-\beta(E_i-E_{i'})}$, where $E_i$ denotes the energy of the $i$-th microstate.
Note that the diagonal terms ensure probability conservation in the system.
It can be shown that $\hat{\mathcal{L}}^\mathrm{cl}$ is Hermitian with respect to the $\boldsymbol{\pi}_\beta$-weighted inner product, 
and it is useful to sort its eigenvalues $\lambda_k$ and its corresponding eigenmodes $\mathbf{r}_k$ as $0 =\lambda_1  > \lambda_2 >\cdots$. 
%
Following Ref.~\cite{Lu2017}, we study a classical spin-$1/2$ chain, where an upwards and a downwards pointing spin on site $n$ is described by $s_n=+1$ and $s_n=-1$, respectively. 
%
%
We study $N_s$ spins interacting locally with their nearest neighbors with coupling $J$ and with a homogeneous external magnetic field with interaction strength $h$, for which the energy of a microstate reads
\begin{equation}
    E_{{i}} = -J\sum_{n=1}^{N_s} s_n s_{n+1} -h\sum_{n=1}^{N_s} s_n \,,
    \label{eq:energy:classical:ising}
\end{equation}
where the microstate index $i \in \{0,1, \dots, 2^{N_s}-1\}$ uniquely labels the configuration $(s_1, s_2, \dots , s_{N_s})$ via binary encoding.
We consider fixed boundary conditions $ s_{N_s+1} = s_1 =+1$ and study the Liouvillian dynamics in \cref{eq:classical:liouville:equation} by plugging \cref{eq:energy:classical:ising} into \cref{eq:classical:liouvillian} under the constraint of allowing only transitions between microstates $i$ and $i'$ that differ by a single spin flip.
Different from~\cite{Lu2017}, for a given initial probability distribution $\mathbf{p}(t_0)$, we construct the fastest-thermalizing, unitarily-connected state.
This is the permutation of the initial $\mathbf{p}$ that minimizes the overlap with the slowest-decaying modes of  $\hat{\mathcal{L}}^\mathrm{cl}$.
Since the number of permutations grows factorially with the number of microstates, we resort to a Metropolis algorithm outlined in~\cite{Moroder2024} to minimize the cost function
{\begin{equation}
    C = \sum_{k=2}^{K+1} \abs{\langle{\mathbf{p}, \mathbf{r}_k}\rangle_{\boldsymbol\pi_\beta} }\,,
    \label{eq:cost:function}
\end{equation}
describing the overlap of the initial state with $K<L-1$ slowest-decaying modes, with
\begin{equation}
    \langle{\mathbf{p}, \mathbf{r}_k}\rangle_{\boldsymbol\pi_\beta}=\sum_i \frac{p_i (\mathbf r_k)_i}{(\boldsymbol \pi_\beta)_i}
\end{equation}
being the so\hyp called Markovian inner product~\cite{Manhart2016}.}

In short, at every iteration, we perform stochastic permutations of four randomly selected spins and always accept the new configuration if the cost function $C'$ is decreased. To avoid local minima, if $C'>C$, we still accept the new configuration with a probability $p_{\mathrm{acc}} = \exp(-T_\mathrm{eff}(C'-C))$, where the effective inverse temperature $T_\mathrm{eff}$ is decreased linearly with the number of successful iterations.
{Note that such optimized (non-equilibrium) initializations are significantly more challenging to prepare experimentally, since, unlike thermal states, they require control over individual spins~\cite{Libal2017}.}

For the thermal operation generated by $\hat{\mathcal{L}}^\mathrm{cl}$, the monotone defined in \cref{eq:quantum-relative-entropy:athermality} simplifies to the \gls{KL}-divergence
\begin{equation}
    M(\mathbf{p}(t)) = D(\mathbf{p}(t) \,\|\, \boldsymbol\pi_\beta) = \sum_{i} p_i(t) \ln\left(\frac{p_i(t)}{ (\boldsymbol \pi_\beta)_i}\right) \,,
    \label{eq:KL}
\end{equation}
which measures the divergence of $\mathbf{p}(t)$ with respect to the thermal state.
The state can be decomposed as
\begin{equation}
\label{eq:state:decomposition:classical}
\begin{split}
\mathbf p(t)=\sum_{k\ge1}a_k\,\mathbf r_k\,e^{\lambda_k t} ,\quad
a_k=\langle\mathbf p,\mathbf r_k\rangle_{\boldsymbol\pi_\beta} \,,
\end{split}
\end{equation}
and asymptotically, the behaviour of the KL-divergence will be 
\begin{equation}
\label{eq:asymptotic:KL}
D\bigl(\mathbf p(t)\,\big\|\,\,\boldsymbol\pi_\beta\bigr) \sim\frac12\,\abs{a_2}^2\,e^{2\Re(\lambda_2) t} \,,
\end{equation}
as shown in \cref{app:asymptotic:decay:qre}.
In other words, the \gls{KL}-divergence decays exponentially with a rate being twice the spectral gap $-\Re(\lambda_2)$.
When the initially more resourceful state has zero overlap with the slowest decaying mode, we are guaranteed to find a Mpemba crossing. Instead,  \cref{eq:asymptotic:KL} allows us to assess whether a crossing will happen also in the context of a weak Mpemba effect (i.e. diminished but not vanishing overlap with the slowest decaying mode for the initially more resourceful state).\\

\begin{figure}
    \centering
    \includegraphics[width=\columnwidth]{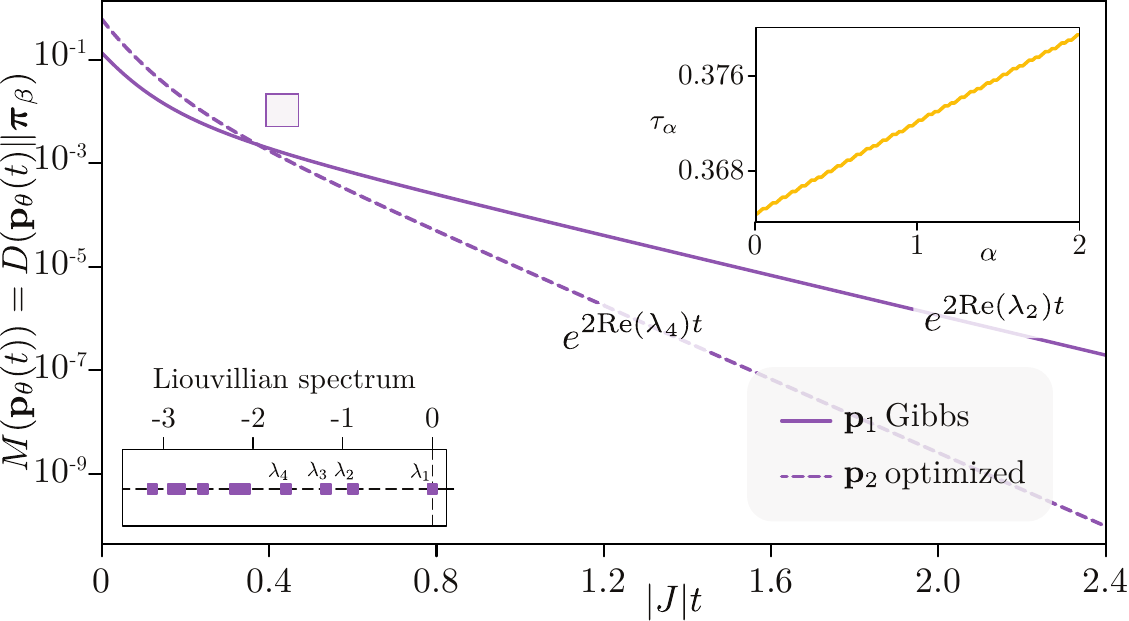}
    \caption{The thermal Mpemba effect in an open classical system \protect\purplesquare\hspace{-0.25em}. For a classical spin chain described by \cref{eq:energy:classical:ising}, we study the \gls{KL}-divergence between the non-equilibrium distribution and the thermal state at $\beta'$ during the thermalization dynamics. Upper-right inset: The crossing times between a thermal and an optimized state for the~$\alpha$-divergence \cref{eq:renyi:divergence} as a function of $\alpha$. Lower-left inset: the ten eigenvalues of the classical Liouvillian $\hat{\mathcal{L}}^\mathrm{cl}$ closest to zero. The second and the fourth eigenvalue determine the asymptotic decay rate of the \gls{KL}-divergence for the thermal and the optimized state, respectively.
    We chose the parameters $J=-0.4$, $h=0.2$, $N_s=7$, the initial {inverse} temperature $\beta' = 0.5$, the bath {inverse} temperature $\beta=1$ {and reduce the cost function~\cref{eq:cost:function} to $C\approx 2\cdot10^{-5}$ for the optimized state}.
    }
    \label{fig:classical:open:KL}
\end{figure}
In \cref{fig:classical:open:KL}, we show the dynamics of the \gls{KL}-divergence during a cooling process for a thermal state {at inverse temperature $\beta'$} (full line) and an optimized state (dashed line) whose overlap with the two slowest-decaying modes of the Liouvillian was minimized.
The optimized state is initially further away from the equilibrium state but thermalizes exponentially faster, displaying an Mpemba crossing~\purplesquare\hspace{-0.25em}.

Similarly to \cref{eq:monotone:quantum:Renyi:divergence:athermality}, the \gls{KL}-divergence can be generalized to a one-parameter family of functions that are also non-increasing under Markovian dynamics, known as the R\'{e}nyi relative entropies~\cite{Tomamichel2016}
\begin{equation}
\begin{split}
M_{\alpha}(\mathbf{p}(t)) =& D_{\alpha} (\mathbf{p}(t) \,\|\, \boldsymbol{\pi}_\beta) \\
=& \frac{1}{\alpha - 1} \ln \left( \sum_i p_i^\alpha(\boldsymbol \pi_\beta)_i^{1-\alpha} \right) \,,
\label{eq:renyi:divergence}
\end{split}
\end{equation}
which correspond to the classical version of \cref{eq:monotone:quantum:Renyi:divergence:athermality}.
The upper-right inset of \cref{fig:classical:open:KL} shows that the Mpemba crossing time $\tau_\alpha$ between the thermal and optimized states increases monotonically with $\alpha$.
This illustrates that $\tau_\alpha$ lacks intrinsic physical significance, as it depends on the choice of divergence measure.
In particular, the variation of $\tau_\alpha$ with $\alpha$ implies that different $\alpha$-divergences can yield qualitatively different conclusions: for certain values of $\alpha$, no crossing may be detected, {as we demonstrate in~\cref{app:alpha:dependent:crossing}. We stress that this argument relies on allowing for general non-equilibrium initial states.}

%
%
\subsection{Example 2: The Davies Map}
%
\label{sec:example:davies:map:thermal}
The generalization of \cref{eq:classical:liouville:equation} to quantum systems is given by the Lindblad equation
\begin{equation}
        \frac{\mathrm{d} \hat{\rho}_s(t)}{\mathrm{d}t} = \mathcal{L} [\hat{\rho}_{s}(t)] \,. 
    \label{eq:Lindblad}
\end{equation}
Here the superoperator $\mathcal{L} = \mathcal{H}_s+ \mathcal{D}$ evolving the state $\hat{\rho}_s$ consists of a unitary part $\mathcal{H}_s\left[\hat{\rho}_s(t)\right] = - i[\hat H_s, \hat{\rho}_s(t)] $, where $\hat H_s$ denotes the system Hamiltonian, and a dissipative part 
\begin{equation}
    \mathcal{D}\left[\hat{\rho}_s(t)\right]=\sum_l \hat{L}_l \hat{\rho}_s(t) \hat{L}_l^\dagger - \frac{1}{2}\left\lbrace\hat{L}_l^\dagger \hat{L}_l, \hat{\rho}_s(t)\right\rbrace \,,
    \label{eq:Dissipator}
\end{equation}
with the jump operators $\hat{L}_l$ describing the effect of an environment on the system.
Here we focus on a Lindbladian describing the weak interaction of a system with a Markovian heat bath, known as the Davies map~\cite{Davies1979}, which is an important example of an elementary Markovian quantum thermal operation~\cite{Lostaglio2018}.
The Davies generator is characterized by two mathematical properties: it satisfies the quantum detailed balance condition~\cite{Alicki1976} and the superoperators corresponding to the unitary and the dissipative part of its generator commute. 
These properties
imply that the thermal (Gibbs) state $\hat\pi_\beta = e^{-\beta\hat H_s}/\Tr[e^{-\beta\hat H_s}]$ is a fixed point of the corresponding Davies map; moreover, under standard irreducibility/ergodicity assumptions (i.e.,
when the semigroup is primitive), this fixed point is unique
(see \cite{Alicki2007}).
Moreover, in the energy eigenbasis, the Davies generator can be written in a block-diagonal form, where one block describes the thermalization of the populations and the other the thermalization of the coherences, which evolve independently from one another (as we will analyze thoroughly in \cref{sec:modes:asymmetry}).
\begin{figure}
    \centering
    \includegraphics[width=\columnwidth]{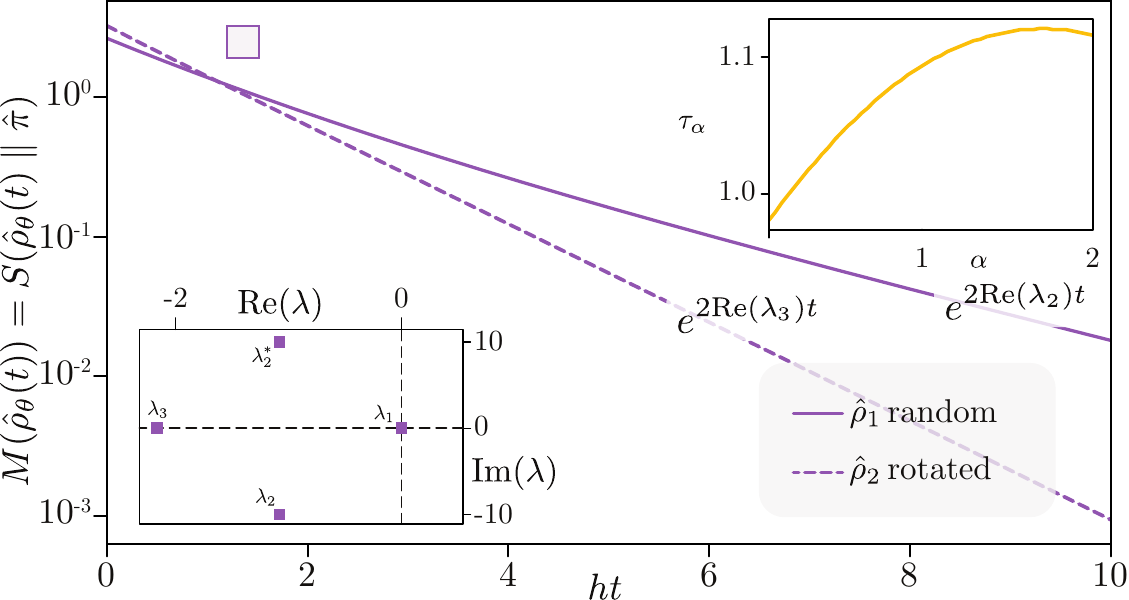}
    \caption{The thermal Mpemba effect in an open quantum system.
    We consider the thermalization of a single qubit described by the Davies map~\cite{Davies1979}. 
     For the state rotated with the unitary transformation outlined in~\cite{Moroder2024} (dashed line), $M(\hat\rho_\theta(t))$ (see \cref{eq:delta:F}) decreases exponentially faster than for a random initial state (full line), displaying an Mpemba crossing.
     Upper-right inset: for the same two states, the Mpemba crossing time $\tau_\alpha$ of the R\'{e}nyi relative entropy $M_\alpha(\hat\rho_\theta(t))$ (\cref{eq:monotone:quantum:Renyi:divergence}) varies with $\alpha$.
     Lower-left inset: different from the classical Liouvillian considered in \cref{fig:classical:open:KL}, the spectrum of the Davies generator also features complex eigenvalues.
    We considered the qubit frequency $h=10$, the bath temperature $\beta=0.1$, a random initial state defined by the Bloch vector $\mathbf{r}= (0.221, 0.867, 0.206)$ and an optimized state with Bloch vector $\mathbf{r}'= (0,0,0.919)$.
    }
    \label{fig:davies:map}
\end{figure}
In~\cite{Moroder2024}, the block structure of the Davies map was exploited to investigate the thermal Mpemba effect.
In particular, it was shown that for any initial state with non-vanishing coherences in the energy eigenbasis, one can always construct an exact unitary transformation $\hat{R}_\mathrm{coh}$ that leads to exponentially faster thermalization, provided that the Liouvillian gap is defined by an eigenvalue with non-vanishing imaginary part, i.e. $\Im(\lambda_2)\neq 0$.
At the same time, a population inversion of the rotated state, implemented by another unitary transformation $\hat{R}_\mathrm{inv}$, always increases the state's non-equilibrium free energy 
\begin{equation}
F(\hat\rho)=\Tr\,\![\hat\rho\, \hat{H}]-\beta^{-1} S(\hat{\rho}) .
\end{equation}
A key aspect of this fundamental thermodynamic quantity is that its (positive) deviation from the equilibrium free energy is proportional to the relative entropy of athermality~\cite{Donald1987}
\begin{equation}
\begin{split}
    \Delta F(t) =& \beta^{-1} M(\hat\rho_s(t)) \\
    =&\beta^{-1}(F(\hat\rho_s(t)) -  F(\hat\pi_\beta)) = \beta^{-1}  S(\hat\rho_s (t)\,\|\, \hat\pi_\beta) \,,
    \label{eq:delta:F}
\end{split}
\end{equation}
where the $M$ is the monotone defined in \cref{eq:quantum-relative-entropy:athermality}.
Crucially, \cref{eq:delta:F} decreases monotonically during the Davies dynamics (see \cref{app:davies:map} for details on the Davies map).\\
By decomposing the state in the eigenmodes of $\mathcal{L}$ similarly to \cref{eq:state:decomposition:classical}, one finds
\begin{equation}
    \hat\rho(t) = \sum_{k\ge1}a_k\,\hat r_k\,e^{\lambda_k t},\quad
a_k=\Tr\,[\hat \ell_k^\dagger \hat \rho] \,,
\end{equation}
which for long times (as we derive in~\cref{app:asymptotic:decay:qre}) implies
\begin{equation}
\label{eq:asymptotic:QRE}
    S(\hat\rho (t)\,\|\, \hat\pi_\beta)\sim \frac{1}{2} \abs{a_j}^2 
 e^{2 \Re(\lambda_j) t} \mathrm{Tr}\bigl[\hat r_j^\dagger \hat \pi^{-1}_\beta \hat r_j\bigr]\,,
\end{equation}
where $j=\argmax_{j:a_j\neq 0}{\Re(\lambda_j)}$ labels the slowest-decaying mode of $\mathcal{L}$ with non-zero overlap with the initial state.
Following~\cite{Moroder2024}, as an example, we consider the Davies map for a single qubit with $\hat H_s = \nicefrac{h}{2}\,\hat\sigma^z$.
In \cref{fig:davies:map}, we study $\Delta F(t)$ for a random initial state (full line) and its corresponding unitarily optimized state (dashed line), which shows the occurrence of a thermal Mpemba effect~\purplesquare\hspace{-0.25em}.
We also consider the monotones defined in   \cref{eq:monotone:quantum:Renyi:divergence:athermality} in the upper-right inset of \cref{fig:davies:map} displaying the Mpemba crossing times as a function of $\alpha$.
As in the classical case discussed in \cref{subs:example:classical:open}, the variation in crossing times $\tau_\alpha$ illustrates that the presence or absence of the Mpemba effect depends on the choice of monotone~\cite{Qian2024,Vanvu2025}, {which we also elaborate on in~\cref{app:alpha:dependent:crossing}}. 
%
Furthermore, the dependence of the Mpemba effect on the monotone function is highlighted in \cref{app:fisher:info}, where we study the Mpemba effect using different \gls{QFI} variants~\cite{Petz1996, Styliaris2020, Scandi2024, Marvian2014}.
Note that in \cref{app:ETH:thermal:Mpemba}, we demonstrate that a quantum thermal Mpemba effect can also arise in a subsystem of an initially pure state evolving under an ergodic Hamiltonian, consistent with the expectations of \gls{ETH}.
\section{The symmetry Mpemba effect}
\label{sec:symmetry:mpemba}
Besides the thermal Mpemba effect, a related phenomenon has recently drawn attention: the anomalous dynamical restoration of symmetries, first investigated by Ares, Murciano, and Calabrese~\cite{Ares2023Nat}.
They considered unitary dynamics generated by a Hamiltonian with a continuous symmetry, where the system is initialized in a state that breaks it.
For a subsystem, the symmetry is (partially) restored during the dynamics, with the remainder of the system acting as a non-Markovian environment.
In this context, the Mpemba effect \protect\orangecircle\ is said to occur when stronger initial symmetry breaking leads to faster symmetry restoration {and the degree of asymmetry is captured by the entanglement asymmetry~\cite{Ares2023Nat}.}
Instances of this effect have been observed in both integrable~\cite{Rylands2024, Murciano2024Stat,DiGiulio2025, Yamashika2024} and non-integrable~\cite{Turkeshi2024, Liu2024, Ares2025randomcircuits} systems, and have been confirmed experimentally~\cite{Joshi2024}. 
{In integrable systems, a compelling physical mechanism underlying the emergence of the Mpemba effect has been identified in Ref.~\cite{Rylands2024}. The key idea is that symmetry restoration is mediated by the spreading of quasiparticles that propagate with different velocities. Consequently, if the dynamics of the initially more asymmetric state involves smaller contributions from the slowest-propagating quasiparticles than those of the initially less asymmetric state, their asymmetry monotones will cross in time. This type of analysis has been applied to the Mpemba effects for the non-Markovian evolution, with a finite-size environment. We therefore believe this applies seamlessly to our non-Markovian circuits of \Cref{sec:U(1),sec:SU(2)}.}

To explain the occurrence of the Mpemba effect in this setting, we first review symmetries in (open) quantum systems in~\cref{sec:symmetries}. We then discuss the resource of asymmetry in~\cref{sec:asymmetry:resource}, and introduce modes of asymmetry, a useful tool for decomposing the dynamics in~\cref{sec:modes:asymmetry}.
{We show the appearance of the effect first using a classical Markov chain in~\cref{sec:example:classical:markov:chain} and its quantum version in~\cref{sec:example:quantum:markov:chain}, then a Davies map in~\cref{sec:example:mpemba:like:open_systems}, and then in scenarios more closely aligned with existing literature: quantum circuits, both interacting with Markovian and with non-Markovian environments with $\mathrm{U}(1)$ symmetry, as well as systems with non-Abelian ($\mathrm{SU}(2)$) symmetries in~\cref{sec:example:mpemba:like:circuits}.}
\subsection{Symmetries in Quantum Systems}
\label{sec:symmetries}
%
%
%
In closed systems, Noether's theorem~\footnote{Noether's first theorem originally applied to continuous (Lie) groups of differentiable symmetries, but its principles have since been extended to include discrete symmetries such as parity or time reversal.} states that each symmetry implies the conservation of a corresponding physical charge~\cite{Kosmann2011,Noether1918}.
When considering isolated quantum systems~\cite{ Albert2014}, the dynamics are governed by a unitary evolution operator $\hat T\in \mathscr{B}(\mathscr{H}_s)$. 
Let $\mathrm{G}$ be a group with unitary representation $\{\hat U_g\}_{g \in \mathrm{G}}\subset \mathscr{B}(\mathscr{H}_s)$.  An operator $\hat T$ is called $\mathrm{G}$\protect\nobreakdash-invariant if 
\begin{equation}
  [\hat T,\hat U_g] = 0 
  \quad\forall\,g\in \mathrm{G}\,,
  \label{eq:G:invariant}
\end{equation}
or, equivalently, $\hat T = \hat U_g^\dagger\,\hat T\,\hat U_g :\forall\,g\in \mathrm{G}$.
%
%
In many cases of interest in physics, the group $\mathrm{G}$ is a continuous one-parameter family represented by unitaries of the form
\begin{equation} 
  \hat U_g = e^{i\,s_g\,\hat Q}\quad :s_g\in \mathbb{R}\,,
  \label{eq:unitary:representation}
\end{equation}
where the $s_g$ provides a group coordinate, while  $\hat Q$ is a Hermitian operator, which will be conserved by all $\mathrm{G}$-invariant unitary transformations, i.e., $\hat{T}\hat Q \hat{T}^\dag=\hat{Q}$ for all $\mathrm{G}$-invariant $\hat{T}$.  This is an example of a connected Lie group. More generally, connected Lie groups of higher dimension, such as $\mathrm{SU}(d)$, have unitary representations given by
\begin{equation}
\hat U_{g}
=
e^{\,i\sum_{\lambda=1}^{\Lambda} s_g^\lambda\,\hat Q_\lambda},\quad \mathbf s_g^\lambda\in \mathbb{R}\,,
\end{equation}
where 
$\{\hat Q_\lambda\},\,\lambda=1,\dots \Lambda$ form a Lie algebra and are, in particular, closed under commutation 
\begin{equation}
[\hat Q_a,\hat Q_b]
=
i\,\sum_{\lambda=1}^{\Lambda}\varepsilon_{ab}^\lambda\,\hat Q_\lambda\,,
\end{equation}
for real $\varepsilon_{ab}^\lambda$. 
%
%
%
%
In spin systems, four paradigmatic continuous symmetries are commonly encountered. \\
(i) Magnetization conservation (phase symmetry, 
    group $\mathrm{U}(1)$):
    \begin{equation}
    \label{eq:unitary:representation:U1}
    \hat{U}_g = e^{i\,\gamma_g\,\hat{S}_z}, \quad \gamma_g \in [0,4\pi)\,,
    \end{equation}
    with $\hat S_z=\sum_n\hat s^z_n$ the total spin-$\tfrac12$ magnetization about the $z$-axis.  In spin-$\tfrac12$ chains, this $\mathrm{U}(1)$ symmetry is formally identical to particle-number conservation in fermionic or hard-core bosonic models.\\
(ii) Total angular momentum conservation, associated
    with the rotation group $\mathrm{SO}(3)$ or its double cover, the group  $\mathrm{SU}(2)$, is
    \begin{equation}
    \hat  U_g
    =e^{i\, \theta_g\, \boldsymbol{\hat{n}}_g\cdot\hat{\mathbf S}}, \quad  \theta_g\in[0,4\pi),\, \boldsymbol{\hat{n}}_g\in\mathbb{R}^3\,,
    \end{equation}
    where $\boldsymbol{\hat{n}}_g\cdot\boldsymbol{\hat{n}}_g=1$
    with $\hat{\mathbf S}=(\hat S_x,\hat S_y,\hat S_z)$, $\hat S_\alpha=\sum_n\hat s^\alpha_n,\, \alpha=x,y,z$.  
  \\
(iii) Energy conservation (time-translation symmetry):
    \begin{equation}
    \hat{U}_g = e^{i\,t_g\,\hat H_s}, \quad t_g \in \mathbb{R}\,.
    \end{equation}
    Here, $\hat H_s$ is the system Hamiltonian, and $t$ represents time, linking this unitary operator to time-translation symmetry. 
    In the special case where the ratios of energy levels are all rational numbers, that is, when the system's time evolution is periodic, this group of unitaries is isomorphic to $\mathrm{U}(1)$. Otherwise, for a non-periodic system, it is isomorphic to $\mathbb{R}$  \footnote{Physically, the more relevant property is that the ratios of Bohr frequencies $E_i - E_j$ are rational numbers, in which case the corresponding unitary group forms a projective representation of $\mathrm{U}(1)$. }.
%

(iv) Another important continuous symmetry, which we
    will not explore further, is that associated with momentum conservation (space-translation symmetry, group $\mathbb{R}^3$):
    \begin{equation}
    \hat{U}_{g} = e^{i\, \mathbf{v}_g\, \cdot \hat{\mathbf{P}}},\quad  \mathbf{v}_g \in \mathbb{R}^3 \, ,
    \end{equation}
    where $\hat{\mathbf{P}} = (\hat{P}_x,\,\hat{P}_y,\,\hat{P}_z)$ is the momentum operator in the three-dimensional space, and $\hat{P}_\alpha = -i\partial_\alpha,\ \alpha = x,y,z$ in position space. 
    Instead, we will consider its discrete version. 
    Consider a 1D equispaced lattice with \gls{PBC} on $N$ sites, the symmetry group is $\mathbb{Z}_N$, realized by the translation operator
    \begin{equation}
    \hat{U}_{g} = e^{i\, {k}_g \,\hat{{P}}}, \quad k_g\in\{0,1,\dots,N-1\} \,,
    \end{equation}
    where $e^{i\hat{{P}}}$ is effectively the space-translation operator with unit lattice constant.
    Recently, the Mpemba effect has been found in this setting for qudits~\cite{Klobas2024}.
    We note that for parity symmetry, associated with the group $\mathbb{Z}_2$, despite a thorough investigation~\cite{Ferro2024,Liu2024}, the Mpemba effect has not been unambiguously observed.

A symmetry group $\mathrm{G}$ acts on density matrices $\hat\rho$ via the unitary representation
\begin{equation}
  \mathcal{U}_g[\hat\rho] = \hat U_g\,\hat\rho\,\hat U_g^\dagger,
  \qquad g\in \mathrm{G}.
\end{equation}
States invariant under every $\mathcal{U}_g$ are called $\mathrm{G}$-invariant.  Given any state $\hat\rho$, its $\mathrm{G}$-invariant component is extracted by the twirling map
\begin{equation}
  \mathcal{G}[\hat\rho]
  = \int_{G}\! \mathrm{d}g\;\mathcal{U}_g[\hat\rho]
  = \int_{G}\! \mathrm{d}g\; \hat U_g\,\hat\rho\,\hat U_g^\dagger,
  \label{eq:twirling}
\end{equation}
where $\mathrm{d}g$ denotes the normalized Haar measure on $\mathrm{G}$, and we have assumed the group $\mathrm{G}$ is compact.  This \gls{CPTP} map projects any input state onto its symmetric component by averaging over all group actions.
For discrete groups this becomes
\begin{equation}
    \mathcal{G}[\hat\rho]
  = \frac{1}{\abs{G}}\sum_{g\in \mathrm{G}} \mathcal{U}_g[\hat\rho]\,.
  \label{eq:twirling:discrete}
\end{equation}

As previously mentioned, the Mpemba effect occurs when the system interacts with an environment. 
{If we therefore consider the $\mathrm{G}$-invariant operator $\hat{T}$ of \cref{eq:G:invariant} as the evolution operator on $\mathscr{H}_s \otimes \mathscr{H}_e$, its local action on $\mathscr{H}_s$ will preserve only a weaker form of symmetry, as follows.}
In fact, let $\mathcal{E}$ be a map defined as 
\begin{equation}\label{eq:dilation}
    \mathcal{E}[\cdot] = \Tr_e[\hat{T}([\cdot]\,\otimes \hat\pi_e)\hat{T}^\dagger]\,,
\end{equation}
with $\hat\pi_e$ a $\mathrm{G}$-invariant state on $\mathscr{H}_e$.
Then, it can be easily shown (see \cref{app:global:invariance:local:covariance}) that, because the initial state of $\mathscr{H}_e$  and the global unitary time evolution $\hat{T}$ respect the symmetry, the time evolution of the reduced system also respects the symmetry; that is, it satisfies the $\mathrm{G}$-covariance condition, also known as weak symmetry~\cite{Buča2012, Albert2014}, namely 
\begin{equation}
\label{eq:covariant:channel}
  \mathcal{E}\circ\mathcal U_g \left[\hat\rho\right]
  =
  \mathcal U_g\circ\mathcal{E}\left[\hat\rho\right],
  \quad\forall\,g\in \mathrm{G},\,\forall\,\hat\rho \,.
\end{equation}
Conversely, it has been shown that any \gls{CPTP} map that respects this covariance condition can be realized via \cref{eq:dilation}, using a $\mathrm{G}$-invariant unitary $\hat{T}$, and a $\mathrm{G}$-invariant state $\hat\pi_e$  \cite{Keyl1999, Marvian_thesis}\footnote{This is sometimes known as covariant Stinespring dilation theorem}.
%
%
%
%
Therefore, while the global evolution is a $\mathrm{G}$-invariant unitary transformation, the dynamics of the reduced state of the system is governed by a $\mathrm{G}$-covariant CPTP map. As a result, the asymmetry (i.e., symmetry breaking) with respect to the group $\mathrm{G}$ can decrease. 
This can be quantified by means of the asymmetry monotones $M$, i.e. functions satisfying 
\begin{equation}
    M\bigl(\mathcal{E}[\hat\rho]\bigr)
  \,\le\,
  M(\hat\rho)\quad\forall\,\hat\rho\in \mathscr{S}(\mathscr{H}_s)
  \label{eq:asymmetry:monotone}
\end{equation}
for any $\mathrm{G}$-covariant $\mathcal{E}$ map.
Hence, while for open systems the presence of symmetry does not imply conservation laws, that is, Noether's theorem is not applicable, asymmetry monotones still capture non-trivial implications of symmetries~\cite{Marvian2014}. 
In generic open system dynamics that respect the symmetry, the asymmetry (symmetry-breaking) in the initial state decreases, although the complete restoration of the symmetry is not a priori guaranteed.
A broad class of maps that are guaranteed to restore a symmetry consists of quantum Markov processes with a fixed point, as shown in Ref.~\cite{Frigerio1978}.
However, non-Markovian examples also exist, such as spin star systems~\cite{Breuer2004}.
In the special case of closed system dynamics, the presence of symmetry implies the conservation of all asymmetry monotones. 
That is, for any $\mathrm{G}$-invariant unitary $\hat{T}$,   $M(\hat{T}\hat\rho\, \hat{T}^\dagger)=M(\hat\rho)$: in this regime, no asymmetry can be created or destroyed.
To observe Mpemba effects, thus one clearly needs to consider $\mathrm{G}$-covariant maps, i.e. systems interacting with environments.\\
%

When $G$ is the one-parameter group of time translations generated by the system Hamiltonian $\hat H_s$ the free operations become the set of $G$-covariant (also called phase-covariant) maps. An example of such maps is the \gls{TO}, defined in \cref{sec:resource_athermality}.
The notion of energetic coherence in quantum thermodynamics, that is, superpositions of states with different energies, can be understood as asymmetry with respect to time translations \cite{Marvian2014b, Lostaglio2015, Marvian2016}. 
Note that although asymmetry under time translations is globally conserved under general energy-conserving unitaries, the reduced map can still dephase in the energy basis, effectively restoring the symmetry by erasing coherence (see \cref{sec:example:mpemba:like:open_systems} for further discussion).
In the open systems literature, the covariance condition is often referred to as weak symmetry, to distinguish it from a stronger notion, known as strong symmetry ~\cite{Buča2012, Albert2014}, in which all the Kraus operators of the channel are $\mathrm{G}$-invariant (i.e., commute with the group action)~\footnote{
    For channels with multiple Kraus operators, the requirement of $\mathrm{G}$-invariance (\cref{eq:G:invariant}) must hold for each Kraus element individually.
    Equivalently, one can demand that the Lindblad generator be unchanged under independent left- and right-unitary twists,
    \begin{equation}
    \label{eq:strong:symmetry}
        \mathcal{U}_{g}^{(L)}\mathcal{L}\,\mathcal{U}_{g}^{(L)\dagger}= \mathcal{U}_{g}^{(R)}\mathcal{L}\,\mathcal{U}_{g}^{(R)\dagger}=\mathcal{L} \quad \forall\,g\in \mathrm{G} \,,
    \end{equation}
    where $\mathcal{U}_{g}^{(L)}\left[\hat\rho\right]=\hat{U}_g\hat\rho$  (ket action) and $\mathcal{U}_{g}^{(R)}\left[\hat\rho\right]=\hat\rho\hat{U}_g^\dagger$  (bra action)~\cite{Buča2012, Albert2014}.
}.
Weak symmetry ($\mathrm{G}$-covariance, \cref{eq:covariant:channel}) admits non-increasing asymmetry monotones, whereas strong symmetry ($\mathrm{G}$-invariance, {\cref{eq:G:invariant}}) guarantees conserved charges in the dynamics.\\
%
%
%
%
Under the restriction to $\mathrm{G}$-covariant operations, any state exhibiting asymmetry cannot be generated for free and must instead be supplied externally as a resource. 
The greater the asymmetry of this resource state, the more power it confers for tasks that require breaking the symmetry.
For instance, in the case of $\mathrm{U}(1)$ symmetry, the precision in probing the phase of a $\mathrm{U}(1)$ rotation increases as the state becomes more tilted with respect to the $z$-direction, as quantified by {\cref{eq:quantum:relative:entropy}}, or, for example, by the Wigner-Yanase skew information (see \cref{app:fisher:info}). Similarly, tasks such as aligning reference frames or achieving quantum-enhanced measurements beyond the shot-noise limit benefit from using more asymmetric states~\cite{Janzing2006,Bartlett2007}.
In this sense, asymmetry becomes a useful resource.
\subsection{Asymmetry as a Resource}
\label{sec:asymmetry:resource}
For any symmetry group $\mathrm{G}$, the resource theory of asymmetry with respect to $\mathrm{G}$ assumes that all systems under consideration are equipped with a representation of the group. 
The set of free states and free operations in this resource theory are then defined as those that respect the symmetry \cite{Marvian_thesis, Gour2008, Marvian2014, Marvian2013}.

In particular, for any system $s$ with Hilbert space $\mathscr{H}_s$, the free states are the $\mathrm{G}$-invariant states,
\begin{equation}
\mathscr{F}
=\left\{\hat\rho \in \mathscr{B}(\mathscr{H}_s) \mid \mathcal{G}\left[\hat\rho\right] = \hat\rho \right\} \,,
\end{equation}
where $\mathcal{G}$ denotes the corresponding twirling map.

Free operations on system $s$ are \gls{CPTP} maps $\mathcal{E}$ satisfying the covariance condition
\begin{equation}
\mathcal{E} \circ \mathcal{U}_g
= \mathcal{U}_g \circ \mathcal{E}
\quad:  \forall\ g \in \mathrm{G}\, .
\tag{\ref{eq:covariant:channel}}
\end{equation}

Note that, as required in all resource theories, free operations (i.e., $\mathrm{G}$-covariant maps) map free states (i.e., $\mathrm{G}$-invariant states) to free states.


%
A useful  monotone in this resource theory is the \emph{relative entropy of asymmetry} \cite{Gour2009, Vaccaro2008}, 
\begin{equation}
\begin{split}
M(\hat\rho)&= \min_{\hat\pi\in\mathscr{F}}
S\bigl(\hat\rho \,\|\, \hat\pi\bigr)  \\  &=S\bigl(\hat\rho \,\|\, \mathcal{G}\left[\hat\rho\right]\bigr)
\\
&= S\bigl(\mathcal{G}\left[\hat\rho\right]\bigr) - S(\hat\rho)\, ,
\label{eq:relative:entropy:asymmetry}
\end{split}
\end{equation}
which corresponds to {\cref{eq:quantum:relative:entropy}}.
This quantifies the minimal divergence of $\hat\rho$ with respect to the manifold of symmetric states. Intuitively, if $\hat\rho$ is nearly invariant, twirling induces little entropy change; if $\hat\rho$ breaks the symmetry strongly, mixing over $\mathrm{G}$ spreads its support and increases entropy substantially~\cite{Marvian2014}.
The relative entropy of asymmetry can be extended to the $\alpha$-divergences of \cref{eq:monotone:quantum:Renyi:divergence} as the $\alpha$-divergences of asymmetry for $\alpha\in(0,1)\cup(1,2]$.
All these maps are guaranteed to be monotonic for $\mathrm{G}$-covariant maps~\cite{Marvian2014}.
An alternative family of monotones are the \gls{QFI}, which we discuss in \cref{app:fisher:info}.

\subsection{Modes of Asymmetry}
\label{sec:modes:asymmetry}

Within the resource theory of asymmetry, an important tool to characterize the symmetry breaking was introduced in Ref.~\cite{Marvian2014b}: the modes of asymmetry.
In this section, we outline the main ingredients to this concept, a simple example, and how this idea can be used to fully understand the symmetry Mpemba effect.

Consider a finite or compact Lie group $\mathrm{G}$ acting unitarily on $\mathscr{H}_s$ via 
$\hat U_g: G \to \mathscr{B}(\mathscr{H}_s)$.  
The full operator algebra $\mathscr{B}(\mathscr{H}_s)$ then carries an induced adjoint action of $\mathrm{G}$, and the irreducible representation of $\mathrm{G}$ labeled by $\mu$, which appears with multiplicity $m_\mu$ in $\mathscr{B}(\mathscr{H}_s)$.  
%
One can use the irreducible tensor operator basis $
  \bigl\{\hat T^{(\mu,q)}_{\alpha}\bigr\}$ of $\mathscr{B}(\mathscr{H}_s)$ $q=1,\dots,d_{\mu}$ ($d_\mu$ dimension of the irrep, namely an irreducible representation) and $\alpha=1,\dots,m_{\mu}$
 satisfying $
  \Tr\bigl[\hat T^{(\mu, q)\dagger}_{\alpha}\,\hat T^{(\nu, \ell)}_{\beta}\bigr]
=\delta_{\mu\nu}\,\delta_{q\ell}\,\delta_{\alpha\beta}, $
%
to decompose any operator $\hat X\in\mathscr{B}(\mathscr{H}_s)$ as
\begin{equation}
  \hat X
  =\sum_{\mu,q}\hat X^{(\mu,q)}.
  \label{eq:generic:mode:decomposition}
\end{equation}
Here, $\hat X^{(\mu,q)}$ is the $(\mu,q)$-mode of $\hat X$, defined as
\begin{equation}
  \hat X^{(\mu,q)}
  =\sum_{\alpha}\Tr\bigl[\hat T^{(\mu,q)}_{\alpha}{}^\dagger\,\hat X\bigr]\;\hat T^{(\mu,q)}_{\alpha}\,.
  \label{eq:generic:mode}
\end{equation}
A basis independent way to define these modes is by the group's action
\begin{equation}
    \hat X^{(\mu,q)}={d}_\mu \int_G \mathrm{d} g\, D_{qq}^{(\mu)*}(g) \,\mathcal{U}_g [\hat X] \,,
  \label{eq:generic:mode:int}
\end{equation}
where $D^{(\mu)}_{\ell q}(g)= \bra{\mu,\ell} \hat U_g\ket{\mu,q}$. 
We write $\mathscr{H}^{(\mu,q)}$ for the subspace where the $(\mu,q)$-mode lives. 
Assuming $\mu=0$ corresponds to the trivial irrep, with $d_\mu=1$, one has $D^{(0)}_{00}(g)=1$ from which it becomes evident how the mode associated with the trivial irrep corresponds to the twirled part of the operator, as in \cref{eq:twirling}.
By orthonormality of the $\bigl\{\hat T^{(\mu,q)}_{\alpha}\bigr\}$ basis it also follows that all the other modes will characterize a symmetry breaking.
For discrete groups \cref{eq:generic:mode:int} becomes
\begin{equation}
    \hat X^{(\mu,q)}=\frac{{d}_\mu}{\abs{G}} \sum_{g\in \mathrm{G}} D_{qq}^{(\mu)*}(g) \,\mathcal{U}_g [\hat X]\,.
  \label{eq:generic:mode:sum}
\end{equation}
This formalism becomes particularly useful when the system evolves by the free operations of the resource theory of asymmetry.
Specifically, following Ref.~\cite{Marvian2014b}, a map $\mathcal{E}$ is $\mathrm{G}$-covariant iff it acts within each $\mu$-sector as
\begin{equation}
  \mathcal{E}\bigl[\hat X^{(\mu,q)}\bigr] =\bigl(\mathcal{E}[\hat X]\bigr)^{(\mu,q)}\,.
  \label{eq:covariance:by:modes}
\end{equation}
%
%
%
%
Namely, each mode evolves independently and so, since the asymmetry can only decrease by $\mathrm{G}$-covariant maps and the trace norm respects the data processing inequality, the trace norm of each asymmetry mode is individually a monotone, that is
\begin{equation}
\,\|\,\mathcal{E}[{\hat X^{(\mu,q)}}]\,\|\,_1 \le\,\|\,{\hat X^{(\mu,q)}}\,\|\,_1\,.
\end{equation}

\begin{figure}
 \centering
 \includegraphics[width=\linewidth]{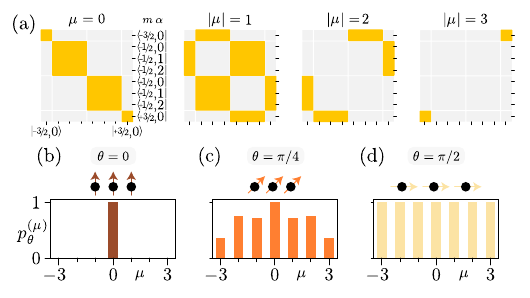}
  \caption{$\mathrm{U}(1)$ symmetry sectors and mode-occupancy distributions for a three-qubit tilted states.
  The computational basis elements labeled as $\ketbra{m,\alpha_1}{m'\!,\alpha_2}$ are associated to a magnetization difference
  $\mu=m-m'$.  
  By reordering the eight basis operators into an $8\times8$ matrix by sorting rows and columns by ascendant $m$, we obtain the $\mathrm{U}(1)$ symmetry sectors shown in panel~(a), with each block (colored in yellow) corresponding to a fixed $\abs{\mu}$.
  Panels~(b–d) display the mode-occupancy histograms (defined in \cref{eq:mode:occupancy}) for the three tilted-ferromagnet states of \cref{eq:tilted:state} with $N_s=3$: in (b) a perfectly symmetric state (only $\mu=0$ appears), in (c) a partially asymmetric state (nonzero weight in several $\mu\neq0$-sectors), and in (d) a maximally asymmetric state (uniform weight across all sectors).}
  \label{fig:tilted-states}
\end{figure}
As a concrete example, consider the symmetry group $G = \mathrm{U}(1)$, generated by the total-$z$ magnetization operator $\hat S_z$ on $N_s$ spin-half particles. 
Its unitary representation is as in \cref{eq:unitary:representation:U1} given by $\hat U_\gamma$ with $\gamma \in [0,4\pi)$. 

Because of the group's simple structure, its asymmetry modes are labeled by a single integer $\mu$,  
which determines the $\mathrm{U}(1)$ irrep as $D^{(\mu)}: e^{i\theta} \mapsto e^{i\mu\theta}$.   
The operator $\hat{S}_z$ is diagonal in the computational basis $\{\ket{0}, \ket{1}\}^{\otimes N_s}$, with eigenvalues  
 $\pm\tfrac{1}{2},\, \dots,\, \pm N_s/2$ when $N_s$, the number of qubits, is odd,  
and $0,\, {\pm} 1 , \dots,\, \pm N_s/2$ when $N_s$ is even.  
We can equivalently label the elements of the computational basis as $\{\ket{m,\alpha}\}$,  
where $m$ is the eigenvalue of $\hat{S}_z$, and $\alpha$ is a multiplicity index.  
In particular, $m$ is simply half the difference between the number of 0's and 1's in a computational basis element. 

Then, the $\mu$-component of any $\hat\rho$ is given by  projection
\begin{equation}
\begin{split}
  \hat\rho^{(\mu)}
  &=
  \int_{0}^{4\pi}\!\frac{\mathrm{d}\gamma}{4\pi}\,
  e^{-i \gamma \mu/2} (e^{i\,\gamma\,\hat{S}_z}\,\hat\rho\,e^{-i\,\gamma\,\hat{S}_z})\\
  &=\sum_m \hat \Pi_m\, \hat \rho \, \hat \Pi_{m+\mu} \,,
\end{split}
\end{equation}
where $\hat \Pi_m=\sum_{\alpha} \ketbra{m,\alpha}{m,\alpha}$ is the projection to the subspace of states with $m$-magnetization. This can also be written as \cref{eq:generic:mode} in terms of the irreducible tensor basis of $\mathrm{U}(1)$ as
\begin{equation}
\hat\rho^{(\mu)}=\sum_\alpha \Tr[\hat T_\alpha^{(\mu)\dagger}\hat \rho]\hat T_\alpha^{(\mu)} \,,
\label{eq:irreducible:basis:U1}
\end{equation}
where one can choose the tensor operators to be simply the outer product of two computational basis elements, or, equivalently, two eigenvectors of $\hat{S}_z$.
\begin{equation}
\hat T_\alpha^{(\mu)}=\ketbra{m,\alpha_1}{m+\mu,\alpha_2}
\end{equation}
where $\alpha=(m,\alpha_1,\alpha_2)$. 
It follows that
\begin{equation}
  e^{i\,\gamma\,\hat{S}_z}\,\hat\rho^{(\mu)}\, e^{-i\,\gamma\,\hat{S}_z}
  = e^{\,i\gamma\mu}\,\hat\rho^{(\mu)}\,,
\end{equation}
where at $\mu=0$ we find the symmetrized (twirled) state.

Fixing the number of qubits to $N_s=3$, we order the basis states by increasing $m_k$ (see panel~(a) of \cref{fig:tilted-states}) to find the block structure with $\mu=0,\pm1,\pm2,\pm3$.
To illustrate the mode decomposition, we employ the family of tilted pure states
\begin{equation}
  \ket{\varphi(\theta)}
  \;=\;
  \bigotimes_{n=1}^{3}\,e^{-i\,\theta\,\hat s^y_n}\,\ket{0}_n,
  \label{eq:tilted:state}
\end{equation}
so that $\theta=0$ gives the fully $z$-aligned (symmetric) state, $\theta=\pi/4$ an intermediate tilt, and $\theta=\pi/2$ the fully $x$-aligned (maximally asymmetric) state (panels~(b-d) of \cref{fig:tilted-states}).  Writing $\hat\rho_\theta = \ketbra{\varphi(\theta)}{\varphi(\theta)}$, we measure the weight of mode $\mu$ via the rescaled trace norm,
\begin{equation}
  p_\theta^{(\mu)}
  \;=\;
  {\,\|\,\hat\rho_\theta^{(\mu)}\,\|_1} / 
       {\,\|\,\hat\rho_{\pi/2}^{(\mu)}\,\|_1}\,.
  \label{eq:mode:occupancy}
\end{equation}
As $\theta$ increases from $0$ to $\pi/2$, the weights $p_\theta^{(\mu\neq0)}$ grow, showing that the tilted state acquires higher-$\mu$ asymmetry components.
%


%
As we discuss in the next section, this formalism helps unify the two phenomena: the thermal Mpemba effect hinges on the small overlap with the overall slowest-decaying eigenmode, whereas the symmetry Mpemba effect stems exclusively from the decay rate of the slowest eigenmode living in a ($\mu\neq0$)-sector.
\subsection{Example 1: A Classical Symmetry Mpemba Effect with 
  \texorpdfstring{$\mathbb{Z}_4$}{Z4}}
\label{sec:example:classical:markov:chain}
\begin{figure}
    \centering
    \includegraphics[width=1\linewidth]{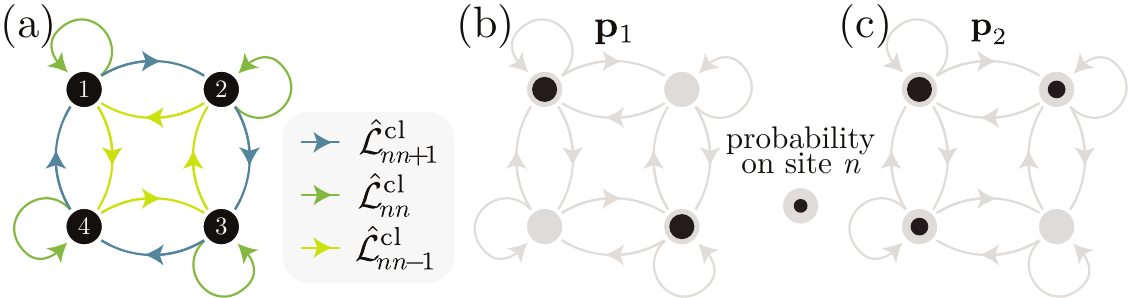}
    \caption{(a) the generator of the classical Markov chain with the four sites arranged on a ring and three types of jumps making them interact. (b) and (c) the distributions $\mathbf{p_1}$ and $\mathbf{p_2}$ which are $(0.5,0,0.5,0)^T$ and $(0.5,0.25,0,0.25)^T$ respectively.}
    \label{fig:classical:symmetric}
\end{figure}

To illustrate how an initially more asymmetric distribution can restore symmetry faster than a more symmetric one \orangecircle, we consider the minimal classical setup in which distinct symmetry-breaking eigenmodes (namely eigenmodes living in $\mu\!\neq\!0$-sectors) decay at different rates: a four-site Markov chain on a ring (represented in~\cref{fig:classical:symmetric}), invariant under the cyclic rotation $n\mapsto n+1\pmod4$, namely $\mathbb{Z}_4$-covariant.
The irreps are $ D^{(\mu)}(n)= e^{2\pi i n \mu / 4}=i^{n \mu}$,  where $\mu=0,1,2,3$.
Jumps occur at a unit rate between each pair of neighbors, as described by the Liouvillian
\begin{equation}
\hat{\mathcal{L}}_{nm}^\mathrm{cl} = 
\begin{cases}
1+\varepsilon,& m = n+1\pmod4\,,\\
1-\varepsilon,& m = n-1\pmod4\,,\\
-2,& n=m\,,\\
0,&\text{otherwise}\,,
\end{cases}
\label{eq:classical:liouvillian:z4}
\end{equation}
where $\varepsilon\in(-1,1)$ is the chirality parameter. 
In this case, there are four modes of asymmetry, corresponding to $\mu=0,1,2,3$. \color{black}
Then, the analog of tensor operators  are given by 
\begin{equation}
\mathbf{T}^{(\mu)}=\tfrac12(1,i^{\mu},i^{2\mu},i^{3\mu})^T\,.
\end{equation}
In particular, under the action of cyclic rotation $n\mapsto n+1\pmod4$, this vector is mapped to itself, up to a phase  
$-i^{\mu}$.
Writing the eigenmodes of $\hat{\mathcal{L}}^\mathrm{cl}$ in terms of the $\{\mathbf{T}^{(\mu)}\}$ basis we get $\boldsymbol{\pi} \! =\! \mathbf{r}_1\! = \! \mathbf{T}^{(0)}$ (eigenvalue $\lambda_1\! =\! 0$), $\mathbf{r}_2 \! =\! \mathbf{T}^{(1)}$ ($\lambda_2 \! =\! -2\!+\!2i\varepsilon$), $\mathbf{r}_3\! =\! \mathbf{T}^{(3)}$ ($\lambda_3\!=\!-2\!-\!2i\varepsilon$), $\mathbf{r}_4 \! =\! \mathbf{T}^{(2)}$ ($\lambda_4\!=\!-4$).

Hence, for $\varepsilon\in(-1,1)$, $\mathbf{r}_4$ is the fastest decaying mode. (Note that the detailed balance condition is violated for $\varepsilon \neq 0$.)
Each eigenmode lives in a single asymmetry sector, which guarantees the preservation of the mode structure, as in \cref{eq:covariance:by:modes}.
Only $\boldsymbol\pi$ (which lives in $\mu=0$) is fully symmetric; the three asymmetric eigenmodes -two slow  \color{black} ones ($\mathbf r_2,\mathbf r_3$) and one fast one ($\mathbf r_4$)- govern how different distortions decay back to equilibrium.
In this case, the relative entropy of asymmetry in \cref{eq:relative:entropy:asymmetry} is equal to the \gls{KL}-divergence of the current distribution $\mathbf{p}(t)$ from its symmetrized version, $\hat{\mathcal{G}}\,\mathbf{p}(t)$, adapted to this case using the discrete version of the twirling operation in \cref{eq:twirling:discrete}.
Since averaging over all four rotations yields the uniform state $  \hat{\mathcal{G}}\,\mathbf{p}(t)=\boldsymbol{\pi}/2$, the relative entropy of asymmetry simplifies to 
\begin{equation}
M\bigl(\mathbf p(t)\bigr)
= D\bigl(\mathbf p(t)\,\|\,\hat{\mathcal G}\,\mathbf p(t)\bigr) = \sum_{k=1}^4 p_k(t)\,\ln\bigl(4\,p_k(t)\bigr)\,,
\end{equation}
which vanishes precisely when $\mathbf p$ is uniform and decreases under the dynamics $e^{t\hat{\mathcal L}^\mathrm{cl}}$.  
To find a strong Mpemba effect, we compare two probability vectors chosen along orthogonal spectral directions as in \cref{eq:state:decomposition:classical}:
\begin{equation}
\mathbf p_1
=\underbrace{\tfrac12\,\boldsymbol\pi}_{\mathbf p_1^{(0)}}
+\underbrace{\tfrac14\,{\mathbf r}_2}_{\mathbf p_1^{(1)}}
+\underbrace{\tfrac14\,{\mathbf r}_3}_{\mathbf p_1^{(3)}}\,,
\quad
\mathbf p_2
=\underbrace{\tfrac12\,\boldsymbol\pi}_{\mathbf p_2^{(0)}}
+\underbrace{\tfrac12\,{\mathbf r}_4}_{\mathbf p_2^{(2)}}\,,
\label{eq:classical:symm:spectral:decomposition}
\end{equation}
where the first lies entirely along the slow  symmetry-breaking eigenmodes $\mathbf r_2$ and $\mathbf r_3$, while the second has only a component in the fast eigenmode $\mathbf r_4$. 
The underbraces highlight the mode decomposition of the states where $\mathbf p_\theta^{(\mu)} = \langle \mathbf T^{(\mu)},\mathbf p_\theta\rangle_{\boldsymbol{\pi}}\, \mathbf T^{(\mu)}$ for $\theta=0,1$ which corresponds to the adaptation of \cref{eq:generic:mode} to this case.
%
%
While at $t=0$, $0.35\approx M(\mathbf p_1)<M(\mathbf p_2)\approx 0.69$, if we pick $\varepsilon=1/4$, according to \cref{eq:asymptotic:KL} their asymmetry behavior asymptotically decreases as 
\begin{equation}
M(\mathbf p_1(t))\sim { \frac14}e^{-4t}\,,
\qquad
M(\mathbf p_2(t))\sim { \frac12}e^{-8t}\,.
\label{eq:asymptotic:relative:entropy:asymmetry}
\end{equation}
By numerically propagating the two initial configurations \cref{eq:classical:symm:spectral:decomposition} with the Liouvillian \cref{eq:classical:liouvillian:z4}, we find that the two monotones $M$ cross at time $\tau = 0.16$.
In other words, the distribution $\mathbf{p}_2$ outpaces $\mathbf{p}_1$ in restoring the $\mathbb{Z}_4$ symmetry, manifesting the symmetry Mpemba effect in a classical system.
In this minimal example, the eigenmodes of the Liouvillian coincide with the irreducible basis, and the uniqueness of the symmetric configuration reduces symmetry restoration to ordinary equilibration on this classical Markov chain.
This simplification, however, is an artifact of our minimal model. 
In the next section, we turn to a richer scenario in which each eigenvalue sector $\mu$ appears with multiplicity 4. 
In that case, one can observe genuinely new behavior, such as nontrivial evolution within the $\mu=0$ symmetric subspace, that cleanly separates the process of symmetry restoration from the usual equilibration dynamics.
\subsection{Example 2: A Quantum Symmetry Mpemba Effect with \texorpdfstring{$\mathbb{Z}_4$}{Z4}}
\label{sec:example:quantum:markov:chain}
In the classical Markov chain of \cref{sec:example:classical:markov:chain}, the generator splits into four one-dimensional symmetry sectors under the action of $\mathbb{Z}_{4}$.
We now construct a quantum analog: a single spinless particle hopping on a four-site ring, governed by a $\mathbb{Z}_{4}$-covariant Lindblad dynamics.
The  irreducible tensor operators are labeled by $\mu\in\{0,1,2,3\}$ with multiplicity index $\alpha\in\{0,1,2,3\}$:
\begin{equation}
  \hat T^{(\mu)}_{\alpha}
  \;=\;\frac{1}{2}\sum_{n=1}^{4}
    e^{{2\pi i n \mu}/{4}}
    \,\ketbra{n+\alpha\hspace{-.9em}\pmod4}{n}
  \,.
\end{equation}
Here $n$ labels the site on the ring, understood modulo 4 so that site 5 is identified with site 1 and site 0 with site 4, and the phase factor ensures each $\hat T^{(\mu)}_{\alpha}$ transforms in the $\mu$-th irreducible representation of $\mathbb{Z}_{4}$.
To define the dynamics, we embed the classical jump-rates into Lindblad hopping operators and add a coherent Hamiltonian term.
The master equation reads
\begin{equation}\label{eq:lindblad:qz4}
\begin{split}
\frac{d\hat\rho}{dt}
\,=&\,\mathcal L[\hat\rho]
\,=\,
-\;i\bigl[\hat H_{s},\,\hat\rho\bigr] + \mathcal{D} _-[\hat\rho] + \mathcal{D}_+[\hat\rho]
\,,
\end{split}
\end{equation}
where the coherent part of the dynamics is generated by the Hamiltonian
\begin{equation}
  \hat H_{s}
  \;=\;
  J\sum_{n=1}^{4}\ketbra{ n}{n+1\hspace{-.9em}\pmod4}
      \;+\;
      \ketbra{n+1\hspace{-.9em}\pmod4}{n}
\end{equation}
which describes hopping between neighboring sites on a four-site ring with amplitude $J$. The incoherent, biased hopping around the ring, is implemented by the two dissipators (defined as \cref{eq:Dissipator}) respectively for the jump operators
\begin{equation}
\begin{aligned}
  \hat L_{n,+}
  &=\sqrt{1+\varepsilon}\; \bigl\lvert\,n+1\hspace{-.9em}\pmod4\bigr\rangle\!\bigl\langle n\bigr\rvert,
  \\
  \hat L_{n,-}
  &=\sqrt{1-\varepsilon}\;\bigl\lvert\,n-1\hspace{-.9em}\pmod4\bigr\rangle\!\bigl\langle n\bigr\rvert\,,
\end{aligned}
\label{eq:jump:operators:z4:quantum}
\end{equation}
where the bias parameter $\varepsilon\in[-1,1]$, as before, tilts the rates so that $\varepsilon>0$ enhances clockwise jumps and $\varepsilon<0$ enhances counterclockwise jumps.
This construction yields a Lindbladian $\mathcal L$ that is manifestly $\mathbb{Z}_{4}$-covariant and thus block-diagonal in the $\mu$-sectors. 

We set $J=1$ and $\varepsilon=1/4$. The Lindblad superoperator $\mathcal L$ admits sixteen eigenmodes, four for each $\mu$-sector, which can be expressed in the tensor basis ${\hat T^{(\mu)}_\alpha}$. Among these, the unique eigenmode belonging to the $\mu\!=\!0$-sector and zero eigenvalue is the steady state $\hat\pi$. The remaining fifteen eigenmodes ${\hat r_k}$ decay with eigenvalues $\lambda_k$, ordered by decreasing real part $\Re(\lambda_k)$~\footnote{In the case of equal real parts, modes are further ordered by increasing $\mu$, and then by decreasing imaginary part.}.

\begin{figure}[t]
\centering
\includegraphics[width=\linewidth]{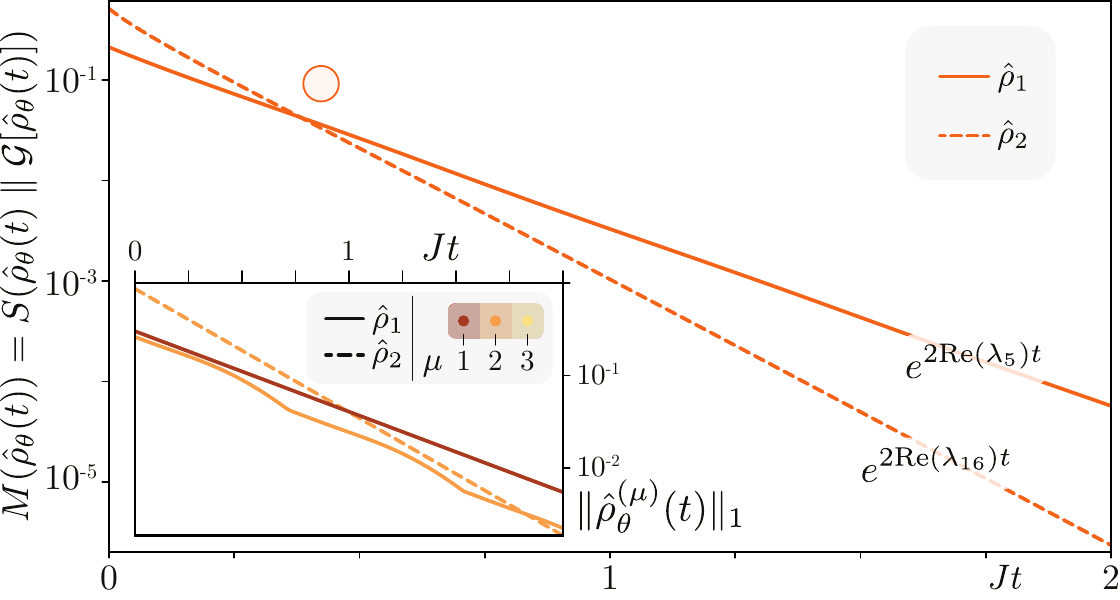}
\caption{Symmetry quantum Mpemba effect in a system with $\mathbb Z_4$ symmetry. 
Main panel: time evolution of the relative entropy of asymmetry $M(\hat\rho_1(t))$ and $M(\hat\rho_2(t))$, showing exponential decay rates given by $\lambda_{10}$ and $\lambda_{16}$, respectively. 
Inset: trace norms $\,\|\,\hat\rho_1^{(\mu)}(t)\,\|\,_1$ and $\,\|\,\hat\rho_2^{(\mu)}(t)\,\|\,_1$ for each charge sector, illustrating that eigenmodes $\hat r_{10}$ and $\hat r_{14}$ are related by Hermitian conjugation and overlap in the $\mu\!=\!1$ and $\mu\!=\!3$ sectors. 
Each eigenmode decays exponentially in time, where the structure of  $\hat \rho_1(t)$ causes $\,\|\,{\hat\rho_1^{(2)}(t)}\,\|\,_1$ to oscillate as it decays.}
\label{fig:symmetric:quantum:z4}
\end{figure}

We prepare two initial states by mixing the steady state with selected decaying eigenmodes:
\begin{equation}
\begin{split}
\hat \rho_1 =& \underbrace{\hat \pi/2 + (\hat r_2 - \hat r_3 + \hat r_4)/9}_{\hat \rho_1^{(0)}} + \underbrace{(-\hat r_{5}/5)}_{\hat \rho_1^{(1)}}\\
&+ \underbrace{(\hat r_{14} + \hat r_{15})/8}_{\hat \rho_1^{(2)}} + \underbrace{(-\hat r_{13}/5)}_{\hat \rho_1^{(3)}}\,,
\\
\hat \rho_2 =& \underbrace{\hat \pi/2 + (\hat r_2  - 2\,\hat r_3  + \hat r_4)/10}_{\hat \rho_2^{(0)}} + \underbrace{-\,\hat r_{16}/2}_{\hat \rho_2^{(2)}}\,.
\end{split}
\end{equation} 

%
These choices introduce different amounts of initial asymmetry in sectors $\mu=1,2,3$. 
We then evolve each state under the dynamics
\begin{equation}
\hat\rho_\theta(t)=e^{t\mathcal L}[\hat\rho_\theta],
\end{equation}
and quantify the restoration of $\mathbb Z_4$ symmetry via the relative entropy of asymmetry $M(\hat\rho_\theta(t))$ as defined in~\cref{eq:relative:entropy:asymmetry}.
\cref{fig:symmetric:quantum:z4} shows that the state $\hat\rho_2$, which starts with a larger asymmetry in sector $\mu=2$, relaxes faster than $\hat\rho_1$, which has its dominant asymmetry in sector $\mu=1$ (and equivalently $\mu=3$). 
Despite having greater initial asymmetry, $\hat\rho_2$ crosses below $\hat\rho_1$ in $M(t)$, demonstrating the symmetry Mpemba effect.
The inset highlights the decay of each charge-sector norm, confirming the underlying exponential behavior of the asymmetry modes.
In this example we see a first advantage brought by the modes of asymmetry: they help identify the eigenmodes responsible for the symmetry breaking.
\subsection{Example 3: Time-Translation-Symmetry Mpemba Effect in Davies Maps}
\label{sec:example:mpemba:like:open_systems}
\begin{figure}[t]
\centering
\includegraphics[width=\linewidth]{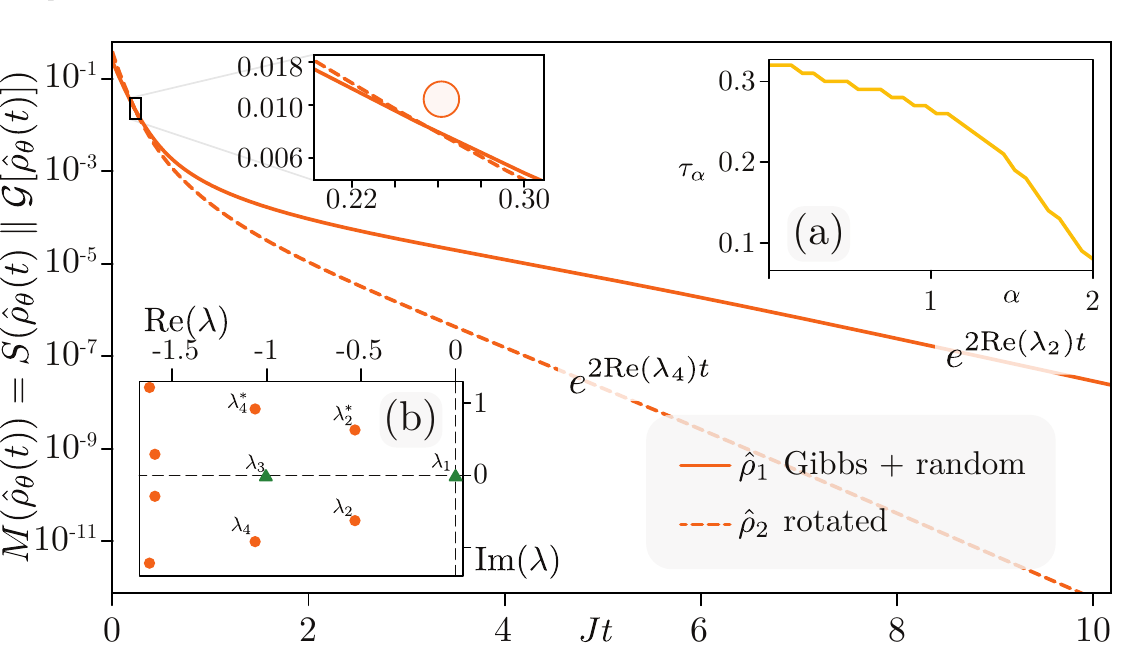}
\caption{Symmetry Mpemba effect in the thermalization of a spin system under the Davies map~\cite{Davies1979}.
We consider a \gls{TFIM} \cref{eq:hamiltonian:ising} with $J=1$, $h=1$, $N_s=4$, environment temperature $\beta=0.1$, initial thermal temperature $\beta_i=1$, and perturbation strength $\gamma=0.05$. 
Note that differently from the previous example \cref{eq:jump:operators:z4:quantum} there is no asymmetry between hopping to the left and to the right (i.e. $\varepsilon=0$).
The solid curve shows $M(\hat\rho_1(t))$ for $\hat\rho_1=(\hat\rho_\mathrm{Gibbs}+\gamma\hat\rho_\mathrm{rand})/N$; the dashed curve shows $M(\hat\rho_2(t))$ for $\hat\rho_2=\hat R_\mathrm{metr}\hat\rho_1 \hat R_\mathrm{metr}^\dagger$, with $\hat R_\mathrm{metr}$ obtained via a Metropolis algorithm \cite{Moroder2024} to minimize overlap with the slowest-decaying eigenmode $\hat \ell_2$. Inset~(a): The Mpemba crossing time for the Petz-R\'{e}nyi $\alpha$-divergences \cref{eq:monotone:quantum:Renyi:divergence} for different $\alpha$. Inset~(b): the ten eigenvalues of the Davies generator with smallest real parts; green triangles indicate ones at $\mu=0$, orange circles the $\mu\neq 0$ ones.}
\label{fig:mpemba_like:davies}
\end{figure}

We now turn to a continuous symmetry, namely the  time-translation symmetry. 
Let us consider once again a Davies map described in \cref{sec:example:davies:map:thermal} but now focusing on the symmetry Mpemba effect~\orangecircle. 
When the symmetry group action ${e^{i\hat H_s t}:t\in\mathbb{R}}$ is generated by the system Hamiltonian $\hat H_s$, the modes of asymmetry $\hat\rho^{(\mu)}$ of a state $\hat\rho$ coincide with its energy coherence blocks at Bohr frequency $\mu$.  Concretely, $\hat\rho=\sum_{\mu}\hat\rho^{(\mu)}$ and $\mathcal{U}_{g}\big[\hat\rho^{(\mu)}\big]\,= e^{i\,\mu \,{{t_g}}}\,\hat\rho^{(\mu)}$, where $\mathcal{U}_{g}[\cdot]=e^{i\hat H_s {t_g}}[\cdot] e^{-i\hat H_s {t_g}}$, so that $\mu=0$ picks out all diagonal (incoherent) terms and each $\mu\neq0$ selects the off-diagonal coherences oscillating at frequency $\mu$ under unitary time evolution. 
The Lindbladian eigenmodes corresponding to symmetry-preserving (symmetry-breaking) modes are depicted in green (orange) in the lower-left inset of \cref{fig:mpemba_like:davies}.
%
%
In Davies maps, the Lindbladian $\mathcal{L}$ commutes by construction with the Hamiltonian evolution and thus respects the time translation symmetry ~\cite{Lostaglio2019}. (Recall that when, up to a shift, the energy levels are all integer multiples of a fixed energy-i.e., the system dynamics is periodic-then time translation becomes isomorphic to $\mathrm{U}(1)$, and such maps are also referred to as phase-covariant.).
As discussed in \cref{sec:symmetries}, this guarantees the monotonicity of the functions
$M_\alpha(\hat\rho(t))$.
%
%
A symmetry Mpemba effect \orangecircle\  occurs when the state with higher initial coherence decoheres faster than one with lower coherence, and their monotone curves cross in time.

As a concrete illustration, consider the transverse-field Ising chain
\begin{equation}
\hat H_s = -J\sum_{n=1}^{N_s-1}\hat\sigma^z_n\hat\sigma^z_{n+1}
+ h\sum_{n=1}^{N_s}\hat\sigma^x_n \,,
\label{eq:hamiltonian:ising}
\end{equation}
with $N_s=4$ and \glspl{OBC}.
We compare two unitarily connected initial states
\begin{align}
\hat\rho_1 &= (\hat\rho_\mathrm{Gibbs} + \gamma\hat\rho_\mathrm{rand})/N \,, \label{eq:thermal:random:state}\\
\hat\rho_2 &= \hat R_\mathrm{metr}\,\hat\rho_1\,\hat R_\mathrm{metr}^\dagger \,,
\end{align}
where $N$ is a normalization constant, $\hat\rho_\mathrm{rand} = \hat X^\dagger\hat X / \Tr\,\! [\hat X]$, with $\hat X$ being a random Hermitian matrix with real and imaginary components uniformly distributed in $[0,1]$ and $\hat R_\mathrm{metr}$ is the unitary transformation minimizing the overlap of $\hat\rho_1$ with $\hat \ell_2$ via the Metropolis algorithm~\cite{Moroder2024}.
\cref{fig:mpemba_like:davies} shows the crossing of $M(\hat\rho_1(t))$ (solid) and $M(\hat\rho_2(t))$ (dashed).
Moreover, the upper-right inset indicates that when considering the Petz-R\'{e}nyi $\alpha$-divergences, the crossing time decreases upon increasing $\alpha$.
Importantly, in contrast to the thermal Mpemba effect (\cref{sec:example:davies:map:thermal}), that involves both $\mu\!=\!0$ (populations) and $\abs{\mu}\!>\!0$ (coherences) blocks, here only the $\abs{\mu}\!>\!0$ modes contribute to the decay of $M(\hat\rho_1(t))$: the slowest-decaying coherent mode corresponds to the complex eigenvalue with real part closest to zero.  
%
%
%
%
%
%
\subsection{Example 4: Symmetry Mpemba Effects in Quantum circuits}
\label{sec:example:mpemba:like:circuits}
\begin{figure}
    \centering
    \includegraphics[width=1\linewidth]{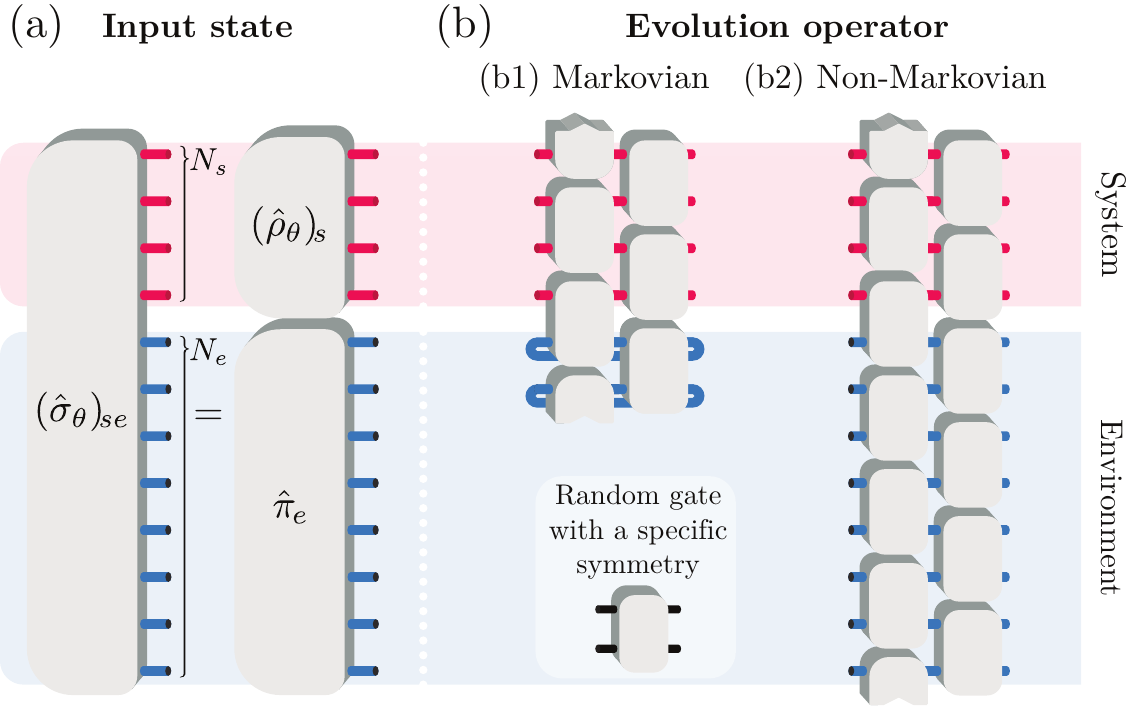}
    \caption{Circuit schematics for the protocols of \cref{sec:example:mpemba:like:circuits}. 
    (a) Preparation of the initial state: the system qubits (pink) are initialized in an asymmetric state parametrized by $\theta$ and the environment qubits in a symmetric state (blue). 
    (b) Evolution operator: successive layers of two-qubit gates, each drawn at random subject to one of the two symmetry classes ($\mathrm{U}(1)$, $\mathrm{SU}(2)$), are arranged in a brickwork geometry.  
    (b1)~The Markovian channel is realized by resetting all environment qubits to the identity state after each $\hat U$ application. 
    (b2)~The non-Markovian channel is performed with no mid-circuit reset or measurement, allowing information backflow from the environment to influence later evolution of the system.
    } 
    \label{fig:circuit}
\end{figure}

Originally, the symmetry Mpemba effect was investigated in continuous-time models~\cite{Ares2023Nat}.
Here we consider circuit models~\cite{Liu2024, Turkeshi2024}, in which time is treated as a discrete variable, and focus on two symmetries, $\mathrm{U}(1)$ and $\mathrm{SU}(2)$.
We consider $N$ qubits, with $N$ even, where the first $N_s$ qubits define the system and the remaining $N_e$ the environment.
The initial state spontaneously breaks the symmetry on the system qubits only, with the extent of the breaking controlled by a parameter $\theta$. The state is then evolved by a circuit composed by gates preserving the symmetry arranged in a brickwork geometry.
We consider two-qubit gates and the Floquet unitary $\hat{U}$ is
\begin{equation}
\hat U \;=\;
\prod_{n=1}^{ N/2} \hat{u}_{2n-1}
\prod_{n=1}^{ N/2} \hat{u}_{2n}\
\,,
\label{eq:brickwork:circuit}
\end{equation}
where $\hat{u}_n$ are gates on the $n,n+1$ qubits and $\hat{u}_{N}$ connects the last and the first qubits to enforce periodic boundary conditions, as represented in the two-qubit brickwork circuits of \cref{fig:circuit}(b).
While in the absence of symmetries any unitary can be decomposed into two local gates, when enforcing a symmetry this is not guaranteed anymore: a generic $\mathrm{G}$-invariant unitary might not be composed of two-local $\mathrm{G}$-invariant gates~\cite{Marvian2022, Marvian2024}. (For $\mathrm{U}(1)$ and $\mathrm{SU}(2)$, two-local gates suffice for semi-universality, meaning that for states restricted to a single irrep of the symmetry, the locality of gates imposes no additional constraints.)
We initialize the system and environment as
\begin{equation}
    \label{eq:circuits:initial:state}
    (\hat\sigma_\theta)_{se} = (\hat \rho_\theta)_s \otimes \hat \pi_e\,,
\end{equation}
picking states that break the symmetry only on the system qubits, and controlling the degree of symmetry breaking through a tilt parameter $\theta$, while $\hat \pi_e$ is a $\mathrm{G}$-invariant state. This guarantees that the system evolution map
\begin{equation}
\mathcal{E}[\hat \rho_{\theta}] = \operatorname{Tr}_e [\hat U\, \hat\sigma_{\theta}\, \hat U^\dagger]
\label{eq:one:time:step:circuit}
\end{equation}
is a non-unitary \gls{CPTP} which is $\mathrm{G}$\protect\nobreakdash-covariant, namely $\mathcal{E}$ is a free operation for the asymmetry resource (see  \cref{app:global:invariance:local:covariance}). 
At each time step, we compute the residual asymmetry with the relative entropy of asymmetry $M(\hat \rho_\theta(t))$,
 as in \cref{eq:relative:entropy:asymmetry}. 
%
%
%
%
%
%
%
\begin{figure}
    \centering
    \includegraphics[width=1\linewidth]{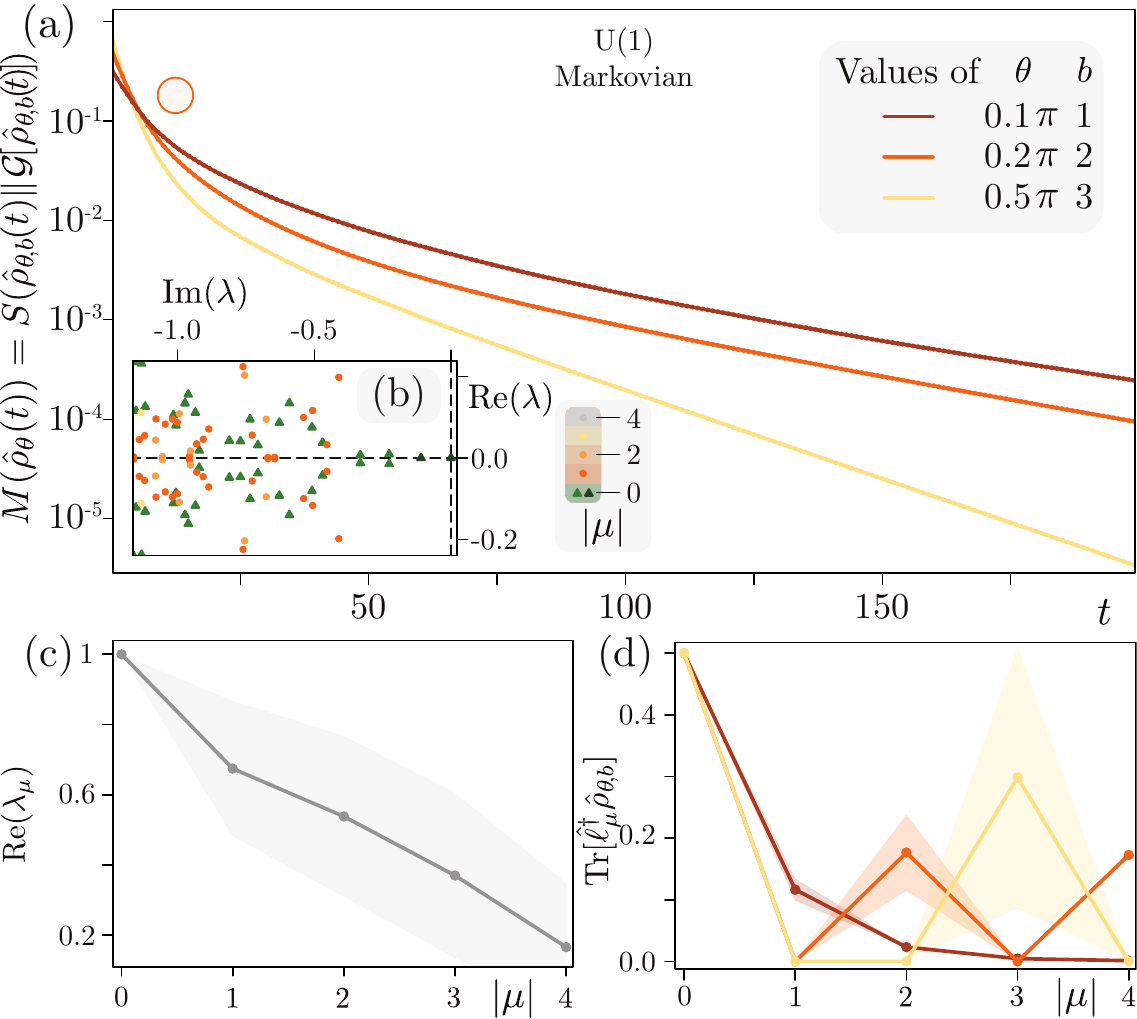}
    \caption{Symmetry-breaking Mpemba effect in a Markovian $\mathrm{U}(1)$ model averaged over 100 circuit realizations with $N_s=4,\ N_e = 2$.
    Panel~(a) shows the time evolution of the monotones.  
    The inset, panel~(b) instead displays the averaged spectrum of the Markovian map, showing how the slow eigenmodes tend to come from low $\abs{\mu}$-sectors, explaining the strong Mpemba effect \protect\orangecircle\ obtained by choosing these three states. 
    Mean (dots) and spread (shaded) of the real part of the slowest decay rate in each $\abs{\mu}$-sector, showing a clear, monotonic decrease with $\abs{\mu}$.
    It is evident that as $\abs{\mu}$ increases, $\Re(\lambda_\mu)$ decreases monotonically.  
    (d) Overlap $\Tr\!\,[\hat\ell_{\mu}^\dagger\,\hat\rho_{\theta,b}]$ of each initial state with its sector's slowest eigenmode $\hat\ell_{\mu}$, confirming that each state couples only to sectors that are multiples of $b$.}
    \label{fig:mpemba:circuits:U1:Markov}
\end{figure}
\subsubsection{Symmetry: U(1)}
\label{sec:U(1)}
We now explore how $\mathrm{U}(1)$ symmetry (i.e. conservation of total  $\hat S^z$) can give rise to a Mpemba effect in random circuits. 
In particular, we show how it is possible to achieve a strong Mpemba effect.
Each two-qubit gate~\cite{Summer2024}
\begin{equation} 
    \hat{u}_{n,n+1} = e^{-i\hat{h}_{n,n+1}}\, e^{-i h_n \hat{s}_n^z}\, e^{-i h'_n \hat{s}_{n+1}^z} \,,
    \label{eq:two-qubit-gate}
\end{equation}
combines an XXZ-type interaction
\begin{equation}
    \hat{h}_{n,n+1} = \frac{J}{2} \Bigl( \hat{s}_n^+ \hat{s}_{n+1}^- e^{i\phi_n} + \hat{s}_n^- \hat{s}_{n+1}^+ e^{-i\phi_n} \Bigr) + J_z\, \hat{s}_n^z \hat{s}_{n+1}^z \,,
    \label{eq:ham-xxz}
\end{equation}
with local $z$-rotations $e^{-i h_n\hat s_n^z}$ and $e^{-i h'n\hat s_{n+1}^z}$. Here $\hat s_n^\pm=\hat s_n^x\pm i\hat s_n^y$, and $\phi_n$ is the Peierls phase. By sampling the five parameters ${h_n,h'_n,\phi_n,J,J_z}$ independently for each gate, the full circuit $\hat U$ remains $\mathrm{U}(1)$-invariant, conserving total magnetization $\hat S^z$.\\

To study the Mpemba effect in circuits we first discuss the clean Markovian limit with a memoryless environment, and then generalize to non-Markovian circuits with a finite number of environment qubits.
In the Markovian limit, the environment is reset after each layer, so that
\begin{equation}
\mathcal{E}_t = (\mathcal{E})^t\,,\qquad \mathcal{E} \equiv\mathcal{E}_1 \,.
\end{equation} 
The environment state $\hat\pi_e$ is the maximally mixed state (see \cref{fig:circuit}(b)), each map $\mathcal{E}$ is $\mathrm{U}(1)$-covariant (which guarantees monotonicity of $M(\hat\rho(t))$) and thus decomposes into asymmetry sectors
\begin{equation}
    \mathcal{E} = \bigoplus_{\mu} \mathcal{E}^{(\mu)}\, 
    \label{eq:decomposition:covariant:map}
\end{equation} 
where $\mathcal{E}^{(\mu)}$ acts on the subspace of the operator space associated with mode $\mu$.
We can diagonalize each $\mathcal{E}^{(\mu)}$ individually using the basis $\{\hat T^{(\mu)}_\alpha\}$ as defined in \cref{eq:irreducible:basis:U1} and obtain the decay eigenmodes in each $\mu$-sector. 

To produce a strong Mpemba effect, we prepare product states of  $N_s$ qubits in blocks of size  $b$ with a uniform tilt $\theta$:
\begin{equation}
\label{eq:state:circuit:U1:system:markovian}
\ket{\varphi(\theta,b)}
=\bigotimes_{n=1}^{\lceil N_s/b\rceil}
\exp\!\Bigl[-\,i\,\theta\;\prod_{j=1}^{b}\hat s^y_{(n-1)b + j}\Bigr]
\;\ket{0}^{\otimes N_s}\,.
\end{equation}
By varying $\theta$ and $b$, we select system states $\hat\rho_{\theta,b} =\ketbra{\varphi(\theta,b)}{\varphi(\theta,b)}$ whose support lies in different $\mu$-sectors (e.g.\ $b=1$ populates all $\mu$-sectors, $b=2$ populates only the even $\mu$-sectors, $b=3$ only sectors multiple of 3, namely $\hat\rho_{\theta,b}=\sum_{k=0}^{\lceil N_s/b\rceil-1}\hat\rho_{\theta,b}^{(kb)}$), so each state $\hat\rho_{\theta,b}$ overlaps differently with the slowest eigenmodes $\{\hat\ell_\mu\}_\mu$ (as in \cref{fig:mpemba:circuits:U1:Markov}(d)) where
\begin{equation}
      \hat\ell_\mu = \hat\ell_{\,k(\mu)}\,,\quad\lambda_\mu = \lambda_{k(\mu)}\,,
\end{equation}
where $k(\mu)$ picks out the slowest eigenmode in the $\mu$-sector
\begin{equation}
      k(\mu) =  \underset{k:\,\hat\ell_k\in \mathscr{H}_s^{(\mu)}}{\arg\max}\, \Re(\lambda_k)\,,
\end{equation}
where $\mathscr{H}_s^{(\mu)}$ is the $\mu$-sector of $\mathscr{H}_s$.
The main panel of \cref{fig:mpemba:circuits:U1:Markov}(a) then reveals distinct asymptotic decay exponents, i.e.\ a strong Mpemba effect.
Avoiding overlap with specific $\mu$–sectors induces a strong Mpemba effect because, in almost every realization of $\hat U$, sectors with larger $\abs{\mu}$ decay more quickly: for $\abs{\mu} > \abs{\mu'}$, one finds $\Re(\lambda_{\mu}) < \Re(\lambda_{\mu'})$.
Indeed, as illustrated in \cref{fig:mpemba:circuits:U1:Markov}, the slowest eigenmodes reside in the blocks with the smallest $\abs{\mu}$ - that is, the lowest absolute $\mathrm{U}(1)$ charge. 
While a more thorough explanation of this phenomenon requires further investigation, one can consider the following possible mechanisms:  
(i) Modes with higher $\mathrm{U}(1)$ charge tend to evolve more rapidly, leading to faster buildup of correlations with the environment and, consequently, quicker information leakage and decay.  
(ii) Since $\mathcal{E}$ is constructed from random $\mathrm{U}(1)$-conserving gates, the blocks corresponding to smaller $\abs{\mu}$, being larger in dimension, are statistically more likely to contain the slowest-decaying eigenmodes.
Moreover, just as in the example of \cref{sec:example:mpemba:like:open_systems}, the initial state's overlap with the slow eigenmodes in sectors with $\mu\neq0$ governs its relaxation rate.
%
Note that the relative entropy of asymmetry $M(\hat \rho(t))$ cannot be interpreted as entanglement asymmetry in this scenario, as the global state is not pure.\\

\begin{figure}
    \centering
    \includegraphics[width=1\linewidth]{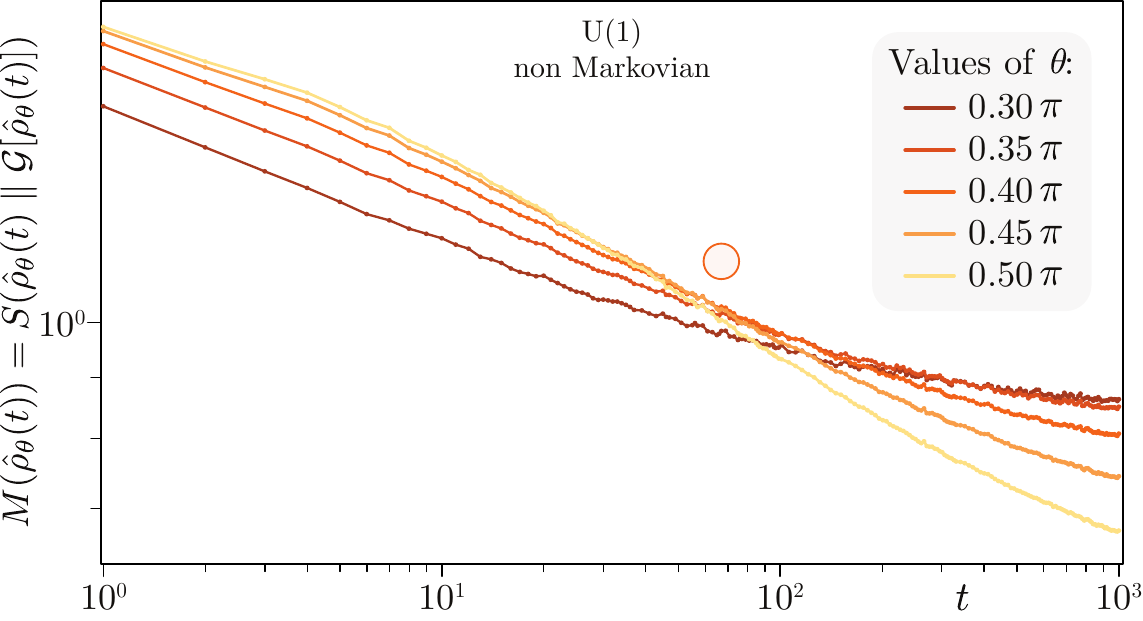}
    \caption{Symmetry-breaking Mpemba effect in a non-Markovian $\mathrm{U}(1)$ model averaged over 100 circuit realizations with $N_s=8,\ N_e = 12$.
    The time evolution of the monotones reveals how the more tilted state, although it starts at a higher amount of asymmetry, restores the symmetry faster. 
    Different from the Markovian case, the monotones do not decay either monotonically or exponentially, as the asymmetry is just distributed in the whole system $+$ environment and therefore finite size effects become more evident.}
    \label{fig:mpemba:circuits:U1:nonMarkov}
\end{figure}
While the Markovian case allows for a clean explanation of the effect, most studies of the symmetry Mpemba effect use non-Markovian circuits. To connect to these, we again prepare a tilted product state on the system (with block size $b=1$) with $\hat \rho_\theta = \ketbra{\varphi(\theta)}{\varphi(\theta)}$ and 
\begin{equation}
\label{eq:state:circuit:U1:system}
\begin{split}
    \ket{\varphi(\theta)} &= \bigotimes_{n=1}^{N_s} e^{-i \theta\,\hat{s}^y_n }\ket{0}_n
\end{split}
\end{equation}
matching Refs.~\cite{Ares2023Nat,Liu2024} except that the tilt now spans only the $N_s$ system qubits, while the environment remains symmetric. Indeed, we initialize the environment in the computational zero state $\hat\pi_e=\ketbra{0}{0}_e=\ketbra{m_\text{min}}{m_\text{min}}_e$, where $\ket{m} 
$ denotes the equal superposition of all basis states with magnetization $m$ and $\hat{s}_n^\alpha$ are the spin-$\tfrac{1}{2}$ operators (with $\alpha\in\{x,y,z\}$).
\cref{eq:decomposition:covariant:map}, which in the Markovian case describes how each symmetry-sector mode decays at a fixed rate, still applies when the map is non-Markovian. However, now it is time-dependent, so both the eigenmodes and the stationary state can shift at each step, while the eigenvalues 
usually shrink exponentially with time when $\hat U$ is ergodic~\cite{Burgarth2013}.
As already discussed, blocks with larger $\abs{\mu}$ admit fewer eigenvalues near unity, so their modes decay more rapidly.
Consequently, an initially highly asymmetric state, which has a larger weight in those fast-decaying, high-frequency sectors, will shed its asymmetry more quickly than a less asymmetric one, yielding the Mpemba effect.\\
%
%
%
\cref{fig:mpemba:circuits:U1:nonMarkov} tracks five such states (different $\theta$) on a system of $N_s=8$ qubits coupled to $N_e=12$ environment qubits. \\

{
Under general covariant time evolutions, asymmetry measures at time $t>0$ are bounded by their value at $t=0$, although in generic non-Markovian processes they can still exhibit transient increases at intermediate times. 
The deviations observed in \cref{fig:mpemba:circuits:U1:nonMarkov} are therefore a consequence of information backflow~\cite{Breuer2016} between the system and environment, amplified by the finite size of the environmental partition. 
We also emphasize that the monotonicity of asymmetry measures follows from the assumption that the environment is initially uncorrelated with the system and prepared in a symmetry-respecting state (see \Cref{app:global:invariance:local:covariance}). 
In contrast, initializing the environment in a symmetry-breaking state, e.g., tilted ferromagnetic states, as done in Refs.~\cite{Liu2024, Turkeshi2024, Ares2025randomcircuits}, violates this condition and can result in a non-$\mathrm G$-covariant reduced map. 
This distinction accounts for the apparent non-monotonic behavior observed in non-Markovian circuit realizations of the $\mathrm{U}(1)$-symmetric Mpemba effect.\\
}

{
In the examples shown in \Cref{fig:mpemba:circuits:U1:Markov}~(c–d), we observed that sectors with larger $|\mu|$ tend to decay faster. 
We expect that a similar phenomenon holds more broadly beyond the Markovian case discussed here, including for random $\mathrm{U}(1)$-invariant circuits with an arbitrary number of layers. 
A better understanding of this phenomenon requires further investigation. 
Our preliminary numerics on such circuits also suggest connections between the quasiparticle picture (Refs.~\cite{Rylands2024,Turkeshi2024}) and the mode-overlap picture, since one can directly compare overlaps with slow quasiparticle carriers and with slow asymmetric eigenmodes.
}

As a final note, these non-Markovian circuits closely mirror setups in deep thermalization with symmetries~\cite{Rui-An2024} and studies of dissipative quantum chaos~\cite{Wold2025}.

\subsubsection{Symmetry: SU(2)}
\label{sec:SU(2)}
\begin{figure}
    \centering
    \includegraphics[width=1\linewidth]{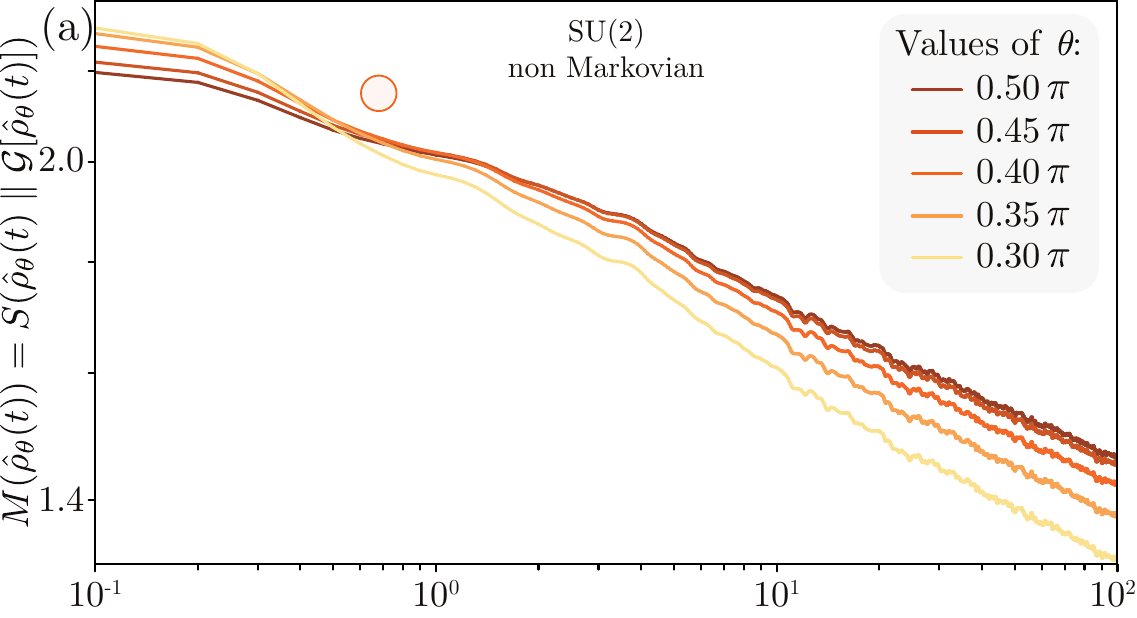}
    \caption{Symmetry Mpemba effect for $\mathrm{SU}(2)$. The initial states are defined in \cref{eq:state:circuit:SU2:system} and we average over $100$ random realizations of the dynamics generated by random $\mathrm{SU}(2)$\hyp symmetric gates with $N_s=8$ and $N_e=12$ drawing $J=J_z$ uniformly from $[-\pi/5,\pi/5]$ for each gate in $\hat U$.}
    \label{fig:SU2}
\end{figure}
We next turn to a non-Abelian example, where the conserved quantity is the total angular momentum $\hat{\mathbf S}^2$. To impose full $\mathrm{SU}(2)$ covariance we switch off all local $Z$-rotations and Peierls phases ($h_n = h'_n = \phi_n = 0$) and sample only the isotropic exchange coupling $J = J_z$.\\
We break the symmetry on the system by preparing the state $\hat \rho_\theta = \ketbra{\varphi(\theta)}{\varphi(\theta)}$ with
\begin{equation}
\ket{\varphi(\theta)}
=
\cos(\theta/2) 
\ket{\xi}^{\otimes \lfloor N_s/2\rfloor} 
+
\sin(\theta/2) \ket{0}^{\otimes N_s}
\,,
\label{eq:state:circuit:SU2:system}
\end{equation}
where $\ket{\xi}=(|01\rangle-|10\rangle)/\sqrt{2}$ is the singlet.
For $\theta \neq k\pi\;\forall\,k\in\mathbb{Z}$, 
we find $\ket{\varphi(\theta)}$ in a superposition of the
$\mathrm{SU}(2)$-invariant singlet $\ket{\xi}^{\otimes\lfloor N_s/2\rfloor}$
and the $\mathrm{SU}(2)$-breaking product $\ket{0}^{\otimes N_s}$.
The environment, which contains an even number of qubits $N_e$, is also initialized in singlets, as $|\xi\rangle^{\otimes N_e/2}$. Note that since the environment's state respects the $\mathrm{SU}(2)$ symmetry, and the Hamiltonian also respects this symmetry, the dynamics of the reduced system likewise respects the symmetry; that is, it is $\mathrm{SU}(2)$-covariant. (When the environment contains an odd number of qubits, it must be prepared in a mixed state in order to obtain an $\mathrm{SU}(2)$-invariant state.)

\cref{fig:SU2} shows five tilted states on an $N_s = 8$ system coupled to $N_e = 12$ environment qubits.
As we vary $\theta$, one can see characteristic crossings in their relaxation curves.
%
%
%
%
%
\section{A unified description of the thermal and the symmetry Mpemba effects}
\label{sec:unified:mpemba}
\begin{figure}
    \centering
    \includegraphics[width=\linewidth]{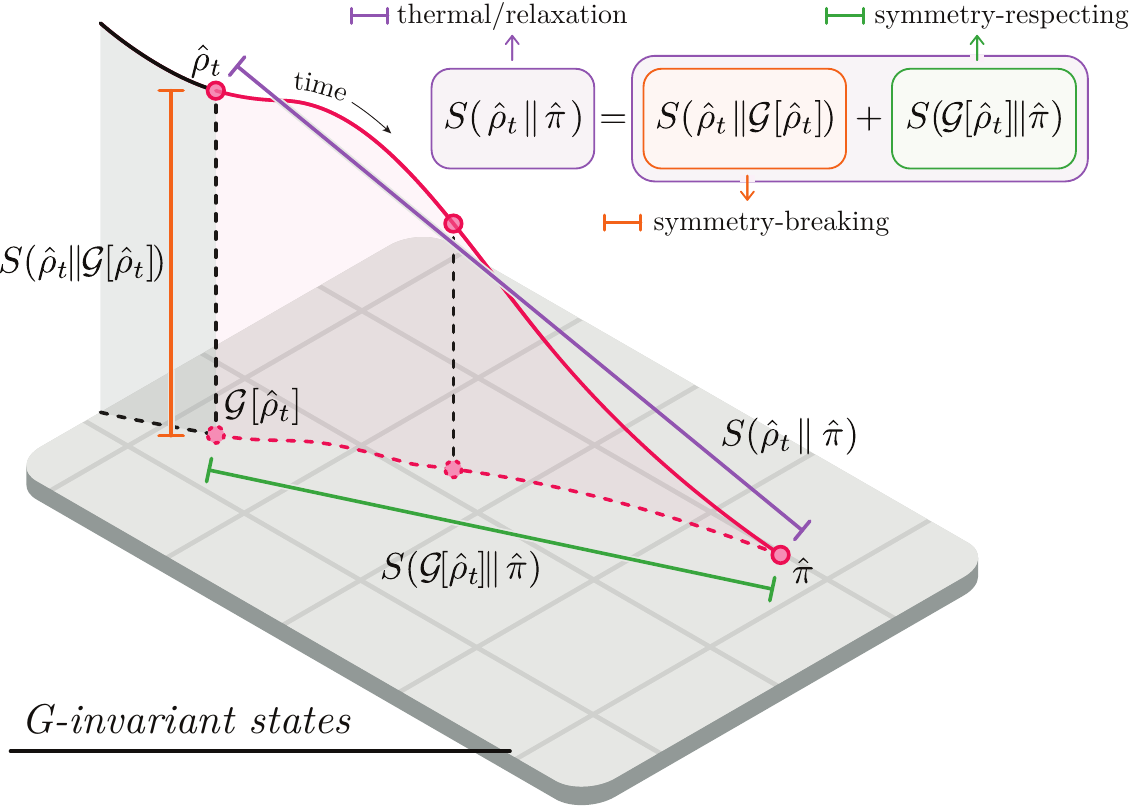}
    \caption{Splitting of the { relative entropy of athermality for systems exhibiting an arbitrary global symmetry into the relative entropy of asymmetry and the relative entropy of athermality for symmetry-respecting states}. 
    An initially asymmetric state $\hat \rho$ evolves under a $\mathrm{G}$-covariant map towards the steady state that lies within the set of $\mathrm{G}$-invariant states (see \cref{sec:symmetries}).
    The relative entropy of { athermality of} $\hat \rho(t)$ ($\hat \rho_t$ in the figure)(violet), can be split into two components: the relative entropy of $\hat \rho(t)$ with respect to the instantaneously dephased state $\mathcal{G}\left[\hat \rho(t)\right]$ (orange, { namely the relative entropy of asymmetry}) and the relative entropy of $\mathcal{G}\left[\hat \rho(t)\right]$ with respect to $\hat \pi$ (green, { namely the relative entropy of athermality for symmetry-respecting states}).   }
    \label{fig:cartoon:relative:entropy:splitting}
\end{figure}
We have reinterpreted instances of both the thermal (\protect\purplesquare \!) and symmetry (\protect\orangecircle) Mpemba effects in systems interacting with featureless baths as well as in non-Markovian environments in a resource theoretical approach in both classical and quantum settings.
Strikingly, the mechanisms behind the thermal and symmetry Mpemba effects are remarkably similar: the former is governed by the slowest-decaying mode, while the latter relies on the slowest symmetry-restoring mode.
%
Importantly, in both cases, the relevant monotone is based on the quantum relative entropy; what differs is the reference state with respect to which it is computed.
For the thermal Mpemba effect~\protect\purplesquare\!\!, it characterizes the divergence with respect to the steady state, while for the symmetry Mpemba effect~\protect\orangecircle, it quantifies the divergence with respect to an instantaneously dephased state.
It is well-known that for the Davies map, the relative entropy between a non-equilibrium state and the thermal steady state can be split into a classical and a purely quantum coherent part~\cite{Santos2019}.
In this context, the system's Hamiltonian is the symmetry generator and, correspondingly, the twirling operation consists of dephasing the non-equilibrium state in the energy eigenbasis. Crucially, here we show that this approach can be generalized to an arbitrary symmetry as follows:
Let $\mathrm{G}$ be a group with unitary representation $\{\hat U_g\}_{g\in \mathrm{G}}$ and $\mathcal{G}$ be the corresponding twirling map as in \cref{eq:twirling}.
For two states $\hat\rho,\hat\pi$ with $\mathcal{G}\left[\hat\pi\right]=\hat\pi$, the quantum relative entropy reads
\begin{equation}
\begin{split}
S(\hat\rho\,\|\,\hat\pi)
&= \Tr\bigl[\hat\rho\ln\hat\rho\bigr]
- \Tr\bigl[\hat\rho\ln\hat\pi\bigr] \\
&\quad + \Tr\bigl[\hat\rho\ln\mathcal{G}\left[\hat\rho\right]\bigr]
- \Tr\bigl[\hat\rho\ln\mathcal{G}\left[\hat\rho\right]\bigr] \\
&= S(\hat\rho\,\|\,\mathcal{G}\left[\hat\rho\right])
- \Tr\bigl[\hat\rho\ln\hat\pi\bigr]
+ \Tr\bigl[\hat\rho\ln\mathcal{G}\left[\hat\rho\right]\bigr]
\, .
\end{split}
\end{equation}
Now since $\hat\pi$ is $G$-invariant, we have that $\Tr\bigl[\hat\rho\ln\hat\pi\bigr]
= \Tr\bigl[\hat\rho
\ln\mathcal{G}[\hat\pi]\bigr]$.
$\mathcal{G}\circ\mathcal{G}=\mathcal{G}$ and $\mathcal{G}^\dagger=\mathcal{G}$ imply
\begin{equation}
\Tr\bigl[\hat\rho\ln\mathcal{G}[\hat\pi]\bigr]
=\Tr\bigl[\hat\rho\,\mathcal{G}[\ln(\mathcal{G}[\hat\pi])]\bigr]=\Tr\bigl[\mathcal{G}[\hat\rho]\,\ln\hat\pi\bigr] \,.
\end{equation}
Analogously, we have that $\Tr\bigl[\hat{\rho} \ln \mathcal{G}[\hat{\rho}]\bigr] = \Tr\bigl[\mathcal{G}[\hat{\rho}] \ln \mathcal{G}[\hat{\rho}]\bigr]$.
Hence, we find that
\begin{equation}
\label{eq:quantum:relative:entropy:splitting}
S(\hat\rho\,\|\,\hat\pi)
= S(\hat\rho\,\|\,\mathcal{G}[\hat\rho])
+ S(\mathcal{G}[\hat\rho]\,\|\,\hat\pi) \,.
\end{equation}
This equality holds for an arbitrary $G$-invariant state $\hat\pi$. 
In the special case where $\hat\pi$ is the thermal state, this identity implies that the  relative entropy of athermality can always be separated into the relative entropy of asymmetry (symmetry-breaking contribution) and the relative entropy of athermality for symmetry-respecting states (symmetry-{respecting} contribution), as illustrated in \cref{fig:cartoon:relative:entropy:splitting}. 
We note that, by \cref{eq:delta:F}, in the special case where $\hat\pi = \hat\pi_\beta$ is the thermal state at temperature $\beta^{-1}$, the relative entropy $S(\hat\rho\,\|\,\hat\pi_\beta)$ can be expressed as the difference between the non-equilibrium free energy of $\hat\rho$ and that of $\hat\pi_\beta$.  
Furthermore, by \cref{eq:relative:entropy:asymmetry}, which was shown in \cite{Gour2009}, we have
$S(\mathcal{G}[\hat\rho]\,\|\,\hat\rho) = S(\mathcal{G}[\hat\rho]) - S(\hat\rho).$
Hence, in this special case, \cref{eq:quantum:relative:entropy:splitting} simplifies to
\begin{equation}
\begin{split}
F(\hat{\rho})-F(\hat\pi_\beta)=
&\beta^{-1} \big[S(\mathcal{G}[\hat\rho])-S(\hat\rho)\big]\\ 
&+F(\mathcal{G}[\hat{\rho}])-F(\hat\pi_\beta)\ ,
\end{split}
\end{equation}
which was previously presented in \cite{Vaccaro2008}.

This decomposition reveals that for Davies maps, the quantum thermal Mpemba effect~\protect\purplesquare comprises a purely classical component~\protect\greentriangle, corresponding to the classical thermal Mpemba effect, and a genuinely quantum part, which is associated with the restoration of time translational symmetry generated by $\hat{H}_s$~\protect\orangecircle.
\begin{figure}
    \centering
    \includegraphics[width=\linewidth]{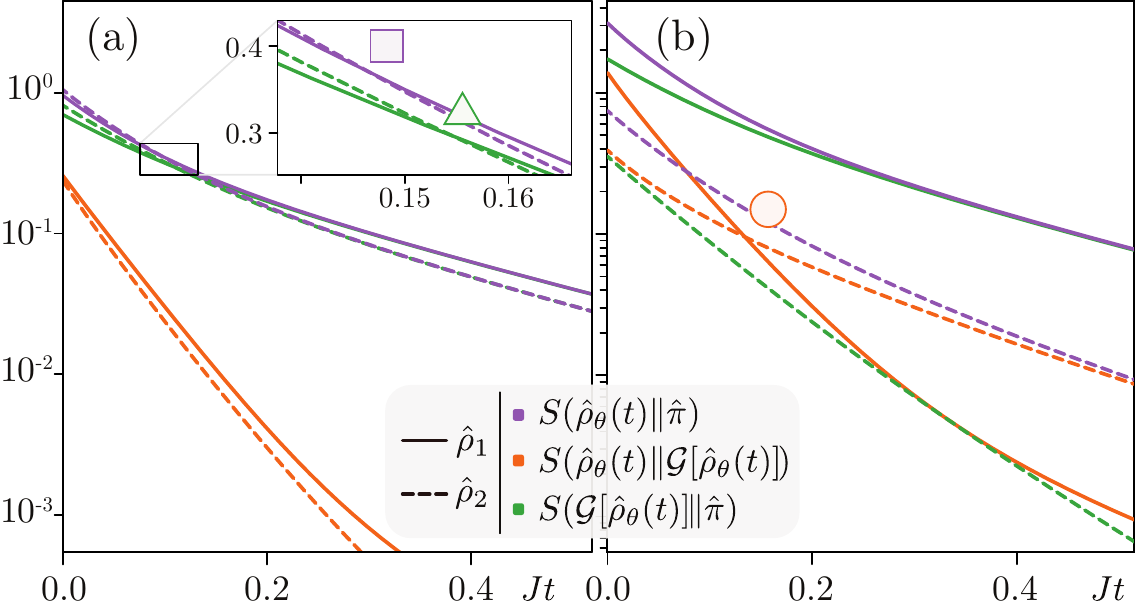}
    
    \caption{Different Mpemba effects in the same thermalization process described by the Davies map. 
    As in \cref{fig:mpemba_like:davies}, we study a \gls{TFIM}, setting $J=1$, $h=1$, $N_s=4$, $\beta=2$, $\beta'=1$ in both panels, and initializing the system in the state $\hat{\rho}_1=\hat{\rho}_\mathrm{Gibbs}^\mathrm{rand} = (\hat{\rho}_\mathrm{Gibbs} + \gamma \hat{\rho}_\mathrm{rand})/N$, and $\hat\rho_2=\hat R_\mathrm{metr}\hat\rho_1 \hat R_\mathrm{metr}^\dagger$ as in \cref{sec:example:mpemba:like:open_systems} (full lines).
    The dashed lines are obtained by stochastically minimizing the overlap with the slowest-decaying mode via the Metropolis algorithm~\cite{Moroder2024}.
    Panel~(a): for weak noise $\gamma=0.05$, the dynamics display a classical thermal \protect\greentriangle\, and a quantum thermal \protect\purplesquare  Mpemba effect.
    Panel~(b): for strong noise $\gamma=0.25$ we find only the symmetry \protect\orangecircle\, Mpemba effect.}
    \label{fig:davies:thermal:and:symmetry}
\end{figure}
Here, as discussed in Ref.~\cite{Moroder2024}, eigenvalues associated with symmetry-preserving eigenmodes are real,  while those corresponding to symmetry-breaking modes are complex (see \cref{fig:mpemba_like:davies}).\ 
We stress that this distinction is not valid for more general G-covariant maps, as shown in  \cref{fig:mpemba:circuits:U1:Markov}.
Moreover, there is nothing intrinsically quantum mechanical about the symmetry-breaking part of the dynamics. In fact, we showed a fully classical instance of a symmetry Mpemba effect in~\cref{sec:example:classical:markov:chain}.
So in general, \cref{eq:quantum:relative:entropy:splitting} decomposes the total athermality (or non-stationarity) monotone into two parts: the asymmetry monotone and a second part that quantifies the symmetry-respecting athermality~\footnote{
The resource theory of symmetry-respecting athermality is defined by the free states 
$\mathcal{F}=\{\hat\rho\mid\mathcal{G}[\hat\rho]=\hat\pi\}$ and the free operations given by $G$‑covariant \glspl{TO}, such as the Davies map in our example. By construction, any free operation $\mathcal{E}$ maps free states to free states: $\mathrm{G}$‑covariance, $\mathcal{U}_g\circ\mathcal{E}=\mathcal{E}\circ\mathcal{U}_g$, implies $\mathcal{G}\circ\mathcal{E}=\mathcal{E}\circ\mathcal{G}$, so for $\hat\rho\in\mathcal{F}$ one has 
\begin{equation}
\mathcal{G}[\mathcal{E}[\hat\rho]]
=\mathcal{E}[\mathcal{G}[\hat\rho]]
=\mathcal{E}[\hat\pi]
=\hat\pi\,,
\end{equation}
hence $\mathcal{E}[\hat\rho]\in\mathcal{F}$.  
}.\\
As a final example, we study the occurrence of these different types of Mpemba effects in the same system.
In \cref{fig:davies:thermal:and:symmetry}, in both panels we consider the same thermalization process described by a Davies map, but for different initializations of the state $\hat{\rho}_1$ as the Gibbs state with random noise defined in \cref{eq:thermal:random:state} with $\gamma=0.05$ (a) and $\gamma=0.25$ (b), while as in \cref{fig:mpemba_like:davies}, $\hat\rho_2=\hat R_\mathrm{metr}\hat\rho_1\hat R_\mathrm{metr}^\dagger$.
Here, the left panel shows both the classical and quantum thermal Mpemba effects, while the right panel displays only the symmetry Mpemba effect. 
While a crossing in either $S(\hat\rho(t)\,\|\,\mathcal{G}[\hat\rho(t)])$ or $S(\mathcal{G}[\hat\rho(t)]\,\|\,\hat\pi)$ is necessary to induce a crossing in the total relative entropy $S(\hat\rho(t)\,\|\,\hat\pi)$, \cref{eq:quantum:relative:entropy:splitting} implies that only when both component entropies cross we are guaranteed to find a crossing in $S(\hat\rho(t)\,\|\,\hat\pi)$.
This example demonstrates that in a given setting, one can seek different types of Mpemba effects, each connected to the depletion of a specific resource, by just changing the resource monotone $M$.\\

%
%
%
%
\section{Conclusion}
\label{sec:conclusion}
We have shown that the various manifestations of the Mpemba effect, arising from the restoration of thermal equilibrium or symmetries, can all be cast within the unified framework of \glspl{RT}.
By formalizing each variant as a distinct resource-theoretic phenomenon, we clarified the conceptual boundaries between thermal Mpemba and symmetry Mpemba effects.
Crucially, we demonstrated that it is not the specific physical embedding of the system (unitary versus featureless bath) that determines which Mpemba effect emerges, but rather the choice of measure used to quantify the system's relaxation.
Indeed, by selecting different resource measures, one can observe multiple Mpemba behaviors in the very same quantum model.
Moreover, due to the absence of total ordering imposed by measures of resource in \glspl{RT}, even when considering the same resource, the existence of the crossing characterizing the Mpemba effect depends on the considered monotone.

We highlighted how the quantum relative entropy has been used in studying the different Mpemba phenomena by taking as reference state the instantaneous projection of the state onto the free state set of the \gls{RT} that corresponds to the Mpemba effect analyzed. 
This reveals a fundamental connection between the thermal Mpemba and the symmetry Mpemba effects.
Crucially, the framework of the modes of asymmetry helps us extend the interpretation of thermal Mpemba effect to the symmetry one by shifting the attention of the initial overlap with the slowest decaying mode to that with the slowest symmetry-breaking one, which dictates the timescale of the symmetry restoration.
Following this approach, we identified the first instance of a symmetry Mpemba effect in a classical system.
Moreover, we analyzed quantum symmetry Mpemba effects for global $\mathrm{U}(1)$ and $\mathrm{SU}(2)$ symmetries, considering both pure and mixed initial states.

The study of Mpemba effects within the framework of \glspl{RT} reveals how different resources can dissipate at different rates, as illustrated in \cref{fig:davies:thermal:and:symmetry}.
{
Furthermore, given a \gls{RT} with a Markovian free‐evolution map $\mathcal{E}$, consider two initial states $\hat\rho_1$ and $\hat\rho_2$ with $M(\hat\rho_1)>M(\hat\rho_2)$.  Let $\hat\ell_j$ be the slowest‐decaying resourceful eigenmode of $\mathcal{E}$ for which $\Tr\!\bigl[\hat\ell_j^\dagger\,\hat\rho_2\bigr]\neq0$.
Then $\hat\rho_1$ and $\hat\rho_2$ exhibit a Mpemba crossing,
\begin{equation}
M(\hat\rho_1(t))<M(\hat\rho_2(t))
\quad\text{for some }t>0\,,
\end{equation}
if
\begin{equation}
\abs{\Tr\!\,[\hat\ell_j^\dagger\,\hat\rho_1]}
\;<\;
\abs{\Tr\!\,[\hat\ell_j^\dagger\,\hat\rho_2]}\,.
\end{equation}
This provides a unified criterion for the occurrence of Mpemba‐type behavior in any \gls{RT}.
}
By analyzing Mpemba dynamics, one can identify the mechanisms that cause certain resources to decay rapidly while others persist. 

%
In this work, we have focused on quantum coherence, athermality, asymmetry, and non-stationarity {(which in the classical context is particularly relevant for driven granular gases~\cite{Lasanta2017, Biswas2020, Teza2025})}, but this approach naturally extends to other resources such as entanglement, magic, and non-Gaussianity.
Beyond unifying thermal and symmetry variants, this framework offers a principled route to designing protocols that accelerate or delay resource dissipation.
Moreover, surveying diverse resource theories may uncover novel Mpemba-type phenomena in both classical and quantum domains.
In this sense, resource theories provide conceptual structure within the ubiquitous but non-universal phenomenon that is the Mpemba effect.
%
%
%

%
%
%
%
\section*{Acknowledgements}
We thank Matteo Lostaglio, Felix Binder, Yutong Luo, Stephen Clark, Oisín Culhane, {and Sara Murciano} for discussions.
This work was supported by the EPSRC-Research Ireland joint project QuamNESS and by Research Ireland under the Frontier For the Future Program. 
J.G. is supported by a Research Ireland Royal Society University Research Fellowship. 
A.S. acknowledges the financial support provided by Microsoft Ireland.
M.M. acknowledges funding from the Royal Society and Research Ireland.
X.T. acknowledges DFG Collaborative Research Center (CRC) 183 Project No. 277101999 - project B01 and DFG under Germany's Excellence Strategy – Cluster of Excellence Matter and Light for Quantum Computing (ML4Q) EXC 2004/1 – 390534769. 
I.M. acknowledges support from 
NSF Phy-2046195, NSF FET-2106448, and 
NSF QLCI grant OMA-2120757.
\section*{Data Availability}
The data that support the findings of this article are openly available at
\url{https://github.com/alessum/resource_theory_mpemba}.
\FloatBarrier
\appendix
\section{Non-Thermal Mpemba Effect and the Resource Theory of non-stationarity}
\label{app:Mpemba:Liouvillian:symmetry}

In the following, we address Mpemba effects in the pure relaxation dynamics of open quantum systems. We consider states $\hat \rho(t)$ evolving under Markovian and time-homogeneous Lindblad dynamics.
We assume that the Liouvillian has a unique, full-rank steady state $\hat{\pi}$ to which any state converges in the long-time limit. The solution to the Lindblad master equation defines a one-parameter semigroup of completely positive trace-preserving (CPTP) maps
\begin{equation}
\mathcal{E}_t = e^{t \mathcal{L}}, \quad t \geq 0 \,.
\end{equation}
We demonstrate that pure relaxation Mpemba effects are captured within the resource theory of non-stationarity, inspired by the resource theory of athermality~\cite{Ng2018}.
The free states in this framework are steady states $\hat{\pi}$
, while any other state is considered resourceful. Free operations $\mathcal{E}$ must satisfy:
\begin{enumerate}
    \item Preservation of the steady state: $\mathcal{E}[\hat{\pi}] = \hat{\pi}$.
    \item Phase covariance: $\mathcal{U}_t \circ \mathcal{E} = \mathcal{E} \circ \mathcal{U}_t$, where $\mathcal{U}_t$ is given by
     \begin{equation}
        \mathcal{U}_t[\cdot] = \hat{\pi}^{it}\,[\cdot]\, \hat{\pi}^{-it} \ , 
    \end{equation}
    for arbitrary $t\in\mathbb{R}$.    
\end{enumerate}
\begin{figure}
    \centering
\includegraphics[width=\linewidth]{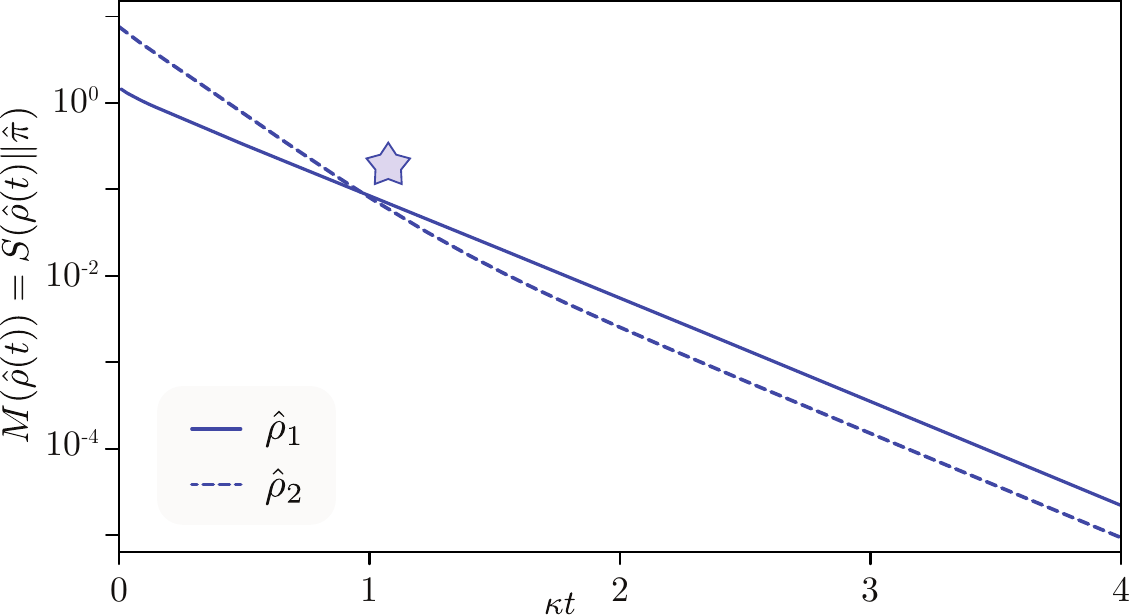}
    \caption{Quantum Mpemba effect in a pure relaxation process of an all-to-all spin model. The monotone quantifying the resource of non-stationarity is the relative entropy between the time-evolved initial states and the fixed point $\hat\pi$.
    Parameters: $N_s = 5$, $\Omega = 1$, $\Delta = -1$, $V=3$, $\kappa = 0.01$. }
    \label{fig:alltoall:model}
\end{figure}
Interestingly, and  generalizing the result on thermalizing maps that connects the athermality monotone to the non-equilibrium free energy, the monotone of non-stationarity,  $S(\hat\rho(t)\,\|\, \hat{\pi})$, is also connected to a known thermodynamic quantity, namely the nonadiabatic entropy production~\cite{Manzano2018} for a full relaxation from $\hat{\rho}(t) \to \hat{\pi}$
\begin{equation}
    \Sigma_\text{na}(t) = M(\hat \rho(t))= S(\hat\rho(t)\,\|\, \hat{\pi}) \,.
\end{equation}
Note that if the fixed point is thermal, the above recovers the total entropy production as originally introduced by Spohn~\cite{Spohn1978}.

As an example, we consider an all-to-all spin model, akin to laser-driven interacting ensembles of Rydberg atoms~\cite{Bloch2008, Wang2024, Lee2012, Kshetrimayum2017}, to demonstrate the Mpemba effect in an open quantum system with a nonthermal fixed point. 
This model has been studied in the context of the quantum Mpemba effect in~\cite{Carollo2021}. 
In adopting their description, we model $N_s$ atoms as two-level systems, described collectively with the spin operators $\hat{S}_\alpha$ and total angular momentum $j^2$. 
For simplicity, we consider the symmetry sector, for which $j^2 = N_s(N_s+2)/4$. 
Its basis is formed by the eigenstates of the $\hat{S}_z$ operator, $\hat{S}_z \ket{m} = m \ket{m}$, with $m$ running from $m=-N_s/2$ to $m=N_s/2$.
The Hamiltonian is given by
\begin{equation}
\label{eq: all:to:all:model}
    \hat{H} = \Omega\, \hat{S}_x - \Delta\, \hat{S}_z + \frac{V}{N_s} \hat{S}_z^2, \quad \hat{L} = \sqrt{\kappa}\,\hat{S}_{-} \,.
\end{equation}
Here, $\hat{S}_{-} = \hat{S}_{x} - i \hat{S}_{y}$, $\Omega$ is the Rabi frequency, $\Delta$ is the laser detuning from the atomic frequency, and $V$ denotes the strength of the all-to-all interactions.
We choose initial states $\hat{\rho}_1$ and $\hat{\rho}_2$ such that $S(\hat\rho_2\,\|\, \hat{\pi})>S(\hat\rho_1\,\|\, \hat{\pi})$. 
As the states evolve under the dynamics generated by the Lindblad master equation specified by the Hamiltonian and jump operator in \cref{eq: all:to:all:model}, we observe a Mpemba effect \protect\bluestar\hspace{-0.25em} in the relative entropy with respect to $\hat \pi$, serving as a monotone for the resource of non-stationarity, shown in \cref{fig:alltoall:model}.
{Finally, we note that the resource of non-stationarity and its associated monotone may also be relevant for certain classical systems. 
In particular, granular gases have been shown to exhibit Mpemba effects, e.g. Refs.~\cite{Lasanta2017, Biswas2020, Teza2025}. 
In regimes where the single particle dynamics are approximately linear and admit a unique steady state, such systems could potentially be described within a similar framework, with the Kullback-Leibler divergence serving as the non-stationarity monotone. 
While extending this perspective to more complex or strongly nonlinear granular gases may not be straightforward, these systems offer an interesting direction for future investigations of Mpemba effects in classical settings.}

\section{Thermal Mpemba Effect with ETH}
\label{app:ETH:thermal:Mpemba}
In the literature, the thermal Mpemba effect has been explored mainly in Markovian open quantum systems obeying \cref{eq:Lindblad}, with the recent exception of~\cite{Bhore2025} where it was also found in isolated systems.
Here we will show that this effect can be found also in a subpart of a quantum system evolving unitarily.
A single qubit, when weakly coupled to a non-integrable environment satisfying the \gls{ETH}~\cite{Srednicki1994, Deutsch1991}, relaxes to a thermal state whose temperature is determined by the total initial energy of the qubit plus environment.
Following ~\cite{ODonovan2024}, we consider the total Hamiltonian
\begin{equation}
    \hat H_\text{se} = \hat H_s + \hat V + \hat H_e \,,
\label{eq:hamiltonian:thermal:mpemba:in:isolated:system}
\end{equation}
with
\begin{equation}
    \hat H_s = h_0\, \hat s_0^z, \hspace{0.7cm} \hat V = \kappa \,\hat s_0^x \otimes \hat s_1^x \,,
\end{equation}
where $\omega_0$ is the system qubit's frequency and $\kappa$ denotes the system-environment coupling.
The environment Hamiltonian is
\begin{equation}
    \hat H_e = J_z \sum_{n=1}^{N-1} \hat s_n^z \hat s_{n+1}^z + h_1 \hat s_1^z+ h_N \hat s_N^z + \sum_{n=1}^{N} h_z \hat s_n^z +  h_x \hat s_n^x.
\end{equation}
%
%
We choose initial states of the form
\begin{equation}
    \ket{\psi_\theta}_{se} = \ket{\varphi_\theta}_s \otimes \ket{E_\theta}_e
\end{equation}
with $\theta=1,2$, $\ket{\varphi_1} = \ket{+}$ and $\ket{\varphi_2}=\ket{0}$.
$\ket{E_\theta}$ is an eigenstate of $\hat{H}_e$ chosen so that the inverse temperature $\beta$ of the overall state (that is fixed by the energy of $\ket{\psi_\theta}_{se}$) matches.
This, together with \gls{ETH}, ensures that the two different initial states on $s$ converge to the same thermal steady state $\hat{\pi}_\beta = e^{-\beta\hat H_s}/\Tr\!\,[-\beta\hat H_s]$, within the fluctuations anticipated by \gls{ETH}.
\\
The evolution of the system's state is described by a \gls{CPTP} map
\begin{equation}
\begin{split}
    \hat{\rho}_\theta(t) =& \mathcal{E}_t[\hat{\rho}_\theta]\\
    =& \Tr_e[e^{-i t\hat H_{se}} (\ketbra{\varphi_\theta}{\varphi_\theta}\otimes \ketbra{E_\theta}{E_\theta})e^{i t\hat H_{se}}]\,.
\end{split}
\end{equation}
To quantify the divergence of $\hat \rho_\theta(t)$ with respect to $\hat \pi_\beta$, we employ the monotone of \cref{eq:quantum-relative-entropy:athermality}. 
As the map is not Markovian, $M(\hat\rho_\theta(t))$ is guaranteed to be monotonic only between time $0$ and $t>0$. At intermediate times, we note small fluctuations which eventually can be considered as finite size effects which would vanish completely when considering an infinitely large bath, for which the map $\mathcal{E}_t$ becomes Markovian ~\cite{Lostaglio2017}.
\begin{figure}
    \centering    \includegraphics[width=1\linewidth]{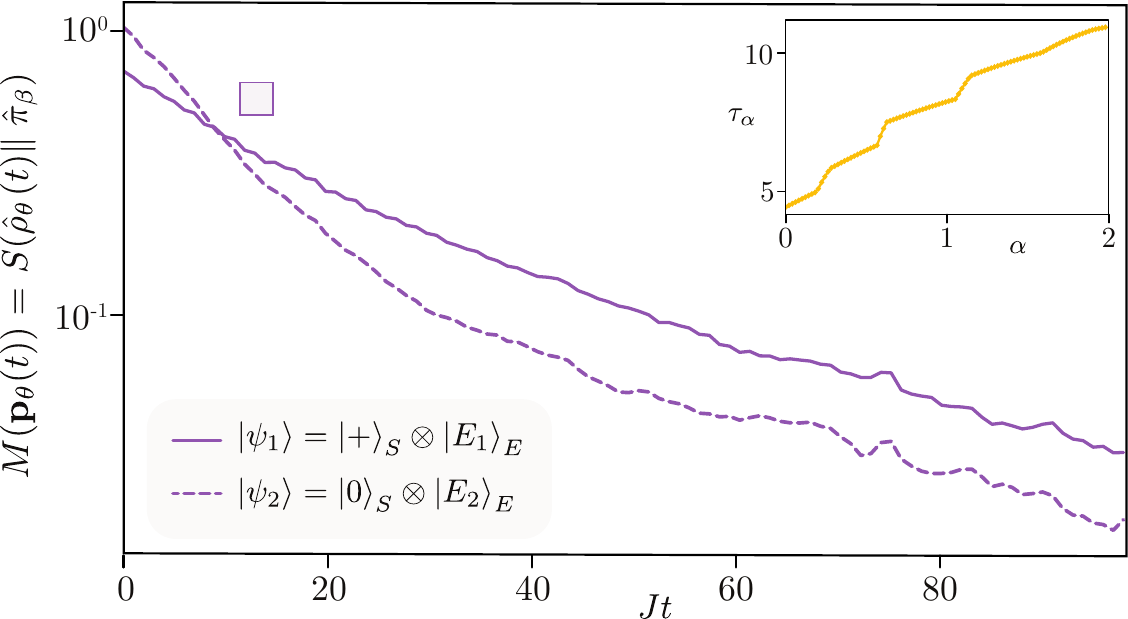}
    \caption{Quantum thermal Mpemba effects in a subpart of a system evolving unitarily. We considered the Hamiltonian \cref{eq:hamiltonian:thermal:mpemba:in:isolated:system} with $(J, h_z, h_x, h_1, h_N, \kappa, h_0) =$ $ (1, 0.3, 1.1, 0.25, -0.25, 0.15, 1.525)$ and $\beta=-0.46$ and $N=15$. Inset: The crossing times $\tau_\alpha$ obtained by studying the different $\alpha$-divergence (see \cref{eq:monotone:quantum:Renyi:divergence}).}
    \label{fig:mpemba:ETH}
\end{figure}
\cref{fig:mpemba:ETH} shows the crossing of the relative entropies indicating the occurrence of a quantum thermal Mpemba effect \protect\purplesquare\hspace{-0.25em}.
The inset shows that  when using $\alpha$-divergences, the crossing time $\tau_\alpha$ depends on $\alpha$.
\section{Asymptotic decay of the quantum relative entropy}
\label{app:asymptotic:decay:qre}
For both classical and quantum systems, the late time behaviour of a state is determined by its overlap with the slowest decaying eigenmode, among the whole spectrum for athermality and only among the nonzero-$\mu$-sectors for asymmetry. Here we prove the equations \cref{eq:asymptotic:KL} and \cref{eq:asymptotic:QRE}.

\subsection{Proof of the Asymptotic Decay of the KL-Divergence}
\label{app:asymptotic:decay:KL}
{\color{black}

Let $\hat{\mathcal{L}}^\mathrm{cl}$ be a classical Liouvillian 
with unique stationary distribution $\boldsymbol\pi$, which we assume has full support ($\pi_i>0$ for all $i$).  
Let us define the weighted inner product 
\begin{equation}
\langle \mathbf u,\mathbf v\rangle_{\boldsymbol\pi}
=\sum_i\frac{u_i\,v_i}{\pi_i}\,,
\end{equation}
and let us order the eigenvalues of $\hat{\mathcal{L}}^\mathrm{cl}$ as
\begin{equation}
0=\Re(\lambda_1)>\Re(\lambda_2)\ge\Re(\lambda_3)\ge\cdots\,.
\end{equation}
We further assume the spectrum to not be degenerate for $\lambda_j$, with $j=2$ in the case of athermality (see~\cref{sec:thermal:mpemba}) and $j$ corresponding to the slowest asymmetric eigenmode for asymmetry (see~\cref{sec:symmetry:mpemba}).
Denoting the corresponding eigenmodes by $\mathbf r_k$ (orthonormal under $\langle\cdot,\cdot\rangle_{\boldsymbol\pi}$),
\begin{equation}
\hat{\mathcal{L}}^\mathrm{cl}\,\mathbf r_k
=\lambda_k\,\mathbf r_k\,,\qquad
\langle \mathbf r_k,\mathbf r_\ell\rangle_{\boldsymbol\pi}
=\delta_{k\ell}\,.
\end{equation}

In this eigenbasis an initial probability distribution $\mathbf p$ evolves as
\begin{equation}
\mathbf p(t)
=\boldsymbol\pi
+\sum_{k\ge2}a_k\,\mathbf r_k\,e^{\lambda_k t},
\quad
a_k=\langle \mathbf p,\mathbf r_k\rangle_{\boldsymbol\pi}\,,
\label{eq:app:classical:time:ev}
\end{equation}
so that $\delta\mathbf p(t)=\mathbf p(t)-\boldsymbol\pi=\sum_{k\ge2}a_k\,\mathbf r_k e^{\lambda_k t}$. 
For small deviations one finds
\begin{equation}
D(\mathbf p(t)\,\|\,\boldsymbol\pi)
=\sum_k p_k(t)\ln\frac{p_k(t)}{\pi_k}
\sim\frac12\sum_k\frac{(\delta p_k(t))^2}{\pi_k}\,.
\end{equation}
As $t\to\infty$, as noted in Ref.~\cite{Polettini2013}, 
only the slowest decaying mode ($j=2$) survives, $\delta p_k\sim a_2\,r_{2,k}\,e^{\lambda_2 t}$, giving
\begin{equation}
D(\mathbf p(t)\,\|\,\boldsymbol\pi)
\sim\frac12\,a_2^2\,e^{2\Re(\lambda_2) t}
\sum_k\frac{r_{2,k}^2}{\pi_k}\,.
\tag{\ref{eq:asymptotic:KL}}
\end{equation}
Replacing $\boldsymbol{\pi}$ with $\hat{\mathcal{G}} \mathbf{p}(t)$ allows to extend this to the asymptotic behaviour of the relative entropy of asymmetry.
}

%
%

\subsection{Proof of the Asymptotic Decay of the Quantum Relative Entropy}
\label{app:asymptotic:decay:quantum}

In both the thermalization example of \cref{sec:example:davies:map:thermal} and the symmetry-breaking scenario of \cref{sec:symmetry:mpemba}, our main monotone
\begin{equation}
M(\hat{\rho}(t))
= S(\hat{\rho}(t)\,\|\, \hat{\pi}_{\beta})
\quad\text{or}\quad
S(\hat{\rho}(t)\,\|\, \mathcal{G}[\hat{\rho}(t)])
\end{equation}
is the quantum relative entropy between the evolving state and its target (the relative entropy of athermality and of asymmetry, respectively). 
{\color{black}
At long times, its decay is governed by the slowest eigenmode of the Lindbladian $\mathcal{L}$ having nonzero overlap with the initial state, following \cref{eq:asymptotic:QRE} for the thermal Mpemba effect and \cref{eq:asymptotic:relative:entropy:asymmetry} (in the case of the $\mathbb{Z}_4$ example studied in \cref{sec:example:classical:markov:chain}) for the asymmetry Mpemba effect.\\

Let now $\mathcal{L}$ be diagonalizable. 
Then there exist right eigenmodes $\{\hat r_k\}$ and left eigenmodes $\{\hat\ell_k\}$ satisfying
\begin{equation}
\mathcal{L}[\hat{r}_k] = \lambda_k\,\hat{r}_k,
\qquad
\mathcal{L}^\dagger[\hat{\ell}_k] = \lambda_k^*\,\hat{\ell}_k,
\end{equation}
with the biorthonormality condition
\begin{equation}
\Tr\!\bigl[\hat{\ell}_k^\dagger\,\hat{r}_\ell\bigr] = \delta_{k\ell},
\end{equation}
and corresponding spectrum ordered so that
\begin{equation}
0 = \Re(\lambda_1) > \Re(\lambda_2) \ge \Re(\lambda_3)\ge\cdots.
\end{equation}
As in \cref{app:asymptotic:decay:KL}, we also assume the slowest eigenmode associated to eigenvalue $\lambda_j$ to be non-degenerate (for athermality $j=1$, for asymmetry the index $j$ is chosen among eigenmodes in $\mu\!\neq\!0$-sectors as in the main text).
}

We now sketch the proof in three steps.
\paragraph{Spectral expansion.}
Analogously to \eqref{eq:app:classical:time:ev}, any initial state decomposes as
\begin{equation}
\hat{\rho}(t) = \hat{\pi} + \sum_{k\ge2} c_k\,\hat{r}_k\,e^{\lambda_k t},
\qquad
c_k = \Tr\!\bigl[\hat{\ell}_k^\dagger\,\hat{\rho}\bigr].
\label{eq:spectral:expansion}
\end{equation}
\paragraph{Quadratic approximation of $S$.}
Writing $\hat{\rho}=\hat{\pi}+\delta\hat{\rho}$ with $\,\|\,\delta\hat{\rho}\,\|\,\ll1$, one shows via the integral representation of the logarithm that
\begin{equation}
S(\hat{\rho}\,\|\,\hat{\pi})
= \Tr\!\bigl[\hat{\rho}(\ln\hat{\rho}-\ln\hat{\pi})\bigr]
\approx \frac12\,
\Tr\!\bigl[\delta\hat{\rho}\,\hat{\pi}^{-1}\,\delta\hat{\rho}\bigr].
\label{eq:second:order:S}
\end{equation}
\paragraph{Dominance of the slowest eigenmode.}
At late times $\delta\hat{\rho}(t)\approx c_j\,\hat{r}_j\,e^{\lambda_j t}$. 
{\color{black}
If $\lambda_j\in\mathbb C\setminus\mathbb R$, its conjugate partner $\lambda_j^*$ must also be included, and one finds
\begin{equation}
\delta\hat{\rho}(t)\approx c_j\,\hat{r}_j\,e^{\lambda_j t}
  + c_j^*\,\hat{r}_j^\dagger\,e^{\lambda_j^* t}.
\end{equation}
Substituting into \eqref{eq:second:order:S} gives
\begin{equation}
\begin{split}
&\Tr\!\bigl[\delta\hat{\rho}\,\hat{\pi}^{-1}\,\delta\hat{\rho}\bigr]\\
&\approx
\begin{cases}
\abs{c_j}^2\,e^{2\Re(\lambda_j) t}\,
\Tr\!\bigl[\hat{r}_j^\dagger\,\hat{\pi}^{-1}\,\hat{r}_j\bigr]\,,
&\lambda_j\in\mathbb R\,,\\ 
2\abs{c_j}^2\,e^{2\Re(\lambda_j) t}\,
\Re(\Tr\![\hat{r}_j^\dagger\,\hat{\pi}^{-1}\,\hat{r}_j])\,,
&\lambda_j\notin\mathbb R\,.
\end{cases}
\end{split}
\end{equation}
Hence,
\begin{equation}
S\bigl(\hat{\rho}(t)\,\|\,\hat{\pi}\bigr)\!
\sim\!
\begin{cases}
\frac12\,\abs{c_j}^2\,e^{2\Re(\lambda_j) t}\,
\Tr\,\!\bigl[\hat{r}_j^\dagger\,\hat{\pi}^{-1}\,\hat{r}_j\bigr],&\lambda_j\in\mathbb R\,,\\
\abs{c_j}^2\,e^{2\Re(\lambda_j) t}\,
\Re(\Tr\,\![\hat{r}_j^\dagger\,\hat{\pi}^{-1}\,\hat{r}_j]),&\lambda_j\notin\mathbb R\,,
\end{cases}
\end{equation}
as discussed in Ref.~\cite{Yu2019}.
Note that this can be straightforwardly extended to $S\bigl(\hat{\rho}(t)\,\|\,\mathcal{G}[\hat{\rho}(t)]\bigr)$ by replacing $\hat\pi$ with $\mathcal{G}[\hat\rho(t)]$ in the proof.
}
\section{{Monotone\hyp dependent occurrence of the Mpemba crossing}}
\label{app:alpha:dependent:crossing}
\begin{figure}
  \centering\includegraphics[width=\linewidth]{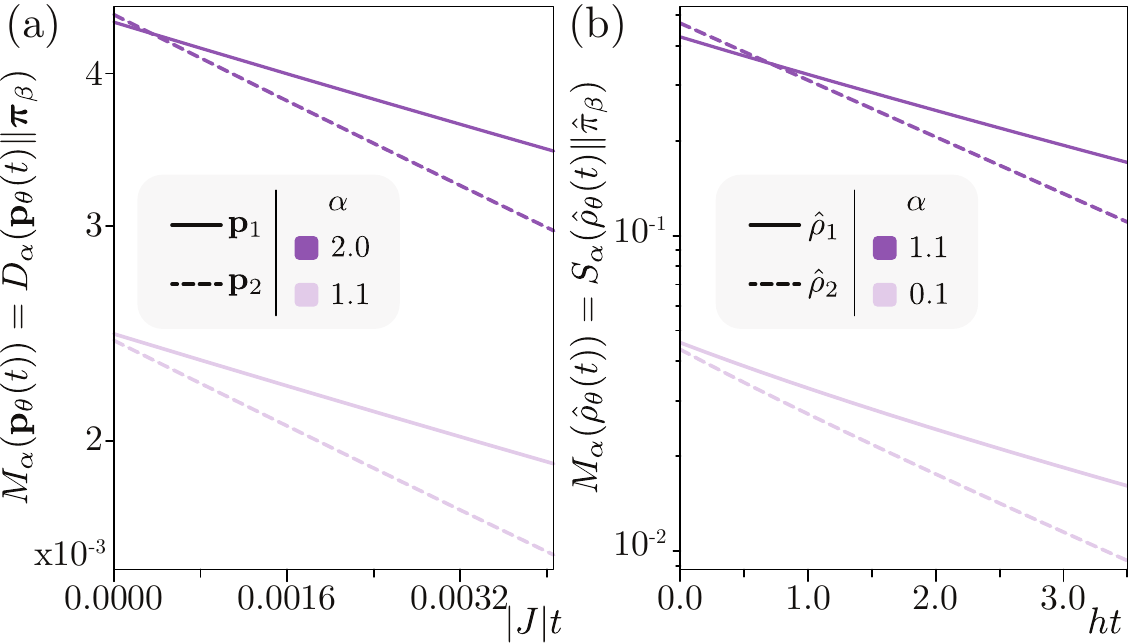}
  \caption{{The presence or absence of the Mpemba crossing depends on the chosen monotone.
  (a) Classical case: we initialize the system with the time-evolved thermal state (full lines) and  optimized state (dashed lines) from \cref{fig:classical:open:KL} at time {$\abs{J}\,t^* \approx 0.376$}. 
  We compute the classical R\'{e}nyi relative entropy~\cref{eq:renyi:divergence} between the time-evolved states and the equilibrium steady state for $\alpha=1.1$ (lighter lines) and $\alpha=2$ (opaque lines).
  In panel (b) we initialize the system with the-time evolved random state (full lines) and optimized state (dashed lines) from \cref{fig:davies:map} at time {$h\,t^* \approx 2$}. 
  We show the quantum R\'{e}nyi relative entropy~\cref{eq:monotone:quantum:Renyi:divergence:athermality} between the time-evolved states and the equilibrium steady state for $\alpha=0.1$ (lighter lines) and $\alpha=1.1$ (opaque lines).
  }}
  \label{fig:app:alpha:dependent:crossing}
\end{figure}
{
In~\cref{fig:classical:open:KL,fig:davies:map} of the main text, we examined the thermal Mpemba effect in a classical and a quantum Markovian setting, respectively. 
The crossing time $\tau_\alpha$, defined as the point where the two $M_\alpha$ curves intersect, was found to vary smoothly with the monotone parameter $\alpha$.
As defined in the main text, these monotones correspond to the classical and quantum R\'{e}nyi relative entropies $M_\alpha$ between the time-evolved states and the equilibrium steady states.
Here, we demonstrate that this dependence of the crossing times on the monotones implies that one can find pairs of states and two parameters $\alpha_1$ and $\alpha_2$ such that a Mpemba crossing appears for $M_{\alpha_1}(t)$ but not for $M_{\alpha_2}(t)$.
This dependence implies that one can find pairs of states and two parameters $\alpha_1$ and $\alpha_2$ such that a Mpemba crossing appears for $M_{\alpha_1}(t)$ but not for $M_{\alpha_2}(t)$.
Since, at least within a certain range, $\tau_\alpha$ increases monotonically with $\alpha$, we can choose the initial states as time-evolved states at a time $t^*$ satisfying $\tau_{\alpha_1} < t^* < \tau_{\alpha_2}$.
Consequently, for $\alpha_1$ the crossing time lies earlier along the forward time axis than $t^*$, whereas for $\alpha_2$ it lies later. 
Because the dynamics are irreversible and backward histories are not unique, ``earlier'' and ``later'' should be understood with respect to forward evolution from the chosen preparatory procedure (the states obtained at $t^*$), rather than as uniquely defined backward trajectories.
}

{
We illustrate this behavior in~\cref{fig:app:alpha:dependent:crossing}, where panel~(a) shows a classical example and panel~(b) shows a quantum one. 
For the classical case, we take as initial states the pair of configurations considered in~\cref{fig:classical:open:KL} at time {$t^* \approx 0.94$}, while for the quantum case we initialize the dynamics with the two states from~\cref{fig:davies:map} at time {$t^* \approx 0.20$}.
This illustrates that, at least when non-equilibrium initial states are allowed, the occurrence of a Mpemba crossing depends on the chosen monotone function used to characterize it, {even when these monotones quantify the same resource} (athermality, in this case).
}

\section{Details on the Davies map}
\label{app:davies:map}
The Davies map arises from a particular Lindblad equation describing the thermalization of quantum systems weakly coupled to a Markovian heat bath \cite{Davies1979}.
In the main text, we have focused on its general mathematical properties, while here we will give the explicit expression of the equation that we used to compute various examples.
Consider the spectral decomposition of the system's Hamiltonian $\hat{H}_s = \sum_k h_k \ketbra{k}{k}$.
In the eigenbasis of $\hat{H}_s$, the Davies generator reads
\begin{equation}
    \begin{split}
    \frac{d\hat{\rho}_s(t)}{dt}=\mathcal{L}\left[\hat{\rho}_s(t)\right] =& -i [\hat{H}_s, \hat{\rho}_s(t)] \\& +\sum_{k<k'}\mathcal{D}^{(1)}_{kk'}[{\hat{\rho}_s(t)}] +\mathcal{D}^{(2)}_{kk'}[{\hat{\rho}_s(t)}],
    \end{split}
    \label{app:eq:davies}
\end{equation}
where the dissipators $\mathcal{D}^{(1)}_{kk'}$ and $\mathcal{D}^{(2)}_{kk'}$ are generated by the jump operators $\hat{L}^{(1)}_{kk'}$ and $\hat{L}^{(2)}_{kk'}$, respectively, which couple all Hamiltonian eigenstates as 
\begin{equation}
\begin{aligned}
        &\hat{L}^{(1)}_{kk'}(\beta) = \left( 1 \mp f^\pm(\beta, h_{k'}\!-\!h_k)\right)^{1/2} \ketbra{k}{k'}\, , \\
        &\hat{L}^{(2)}_{kk'}(\beta) = \left(f^\pm(\beta,h_{k'}\!-\!h_k)\right)^{1/2} \ketbra{k'}\,.
\end{aligned}
    \label{eq:davies:jumpop:fermions}
\end{equation}
Here $f^\pm(\beta, h_{k'}\!-\!h_k) = 1/( \exp(\beta (h_{k'}\!-\!h_k)) \pm 1)$ are the Fermi ($+$) and the Bose ($-$) distribution functions which enforce the detailed balance condition, ensuring that the unique steady state is the thermal state at inverse temperature $\beta$. 
Throughout all examples in the main text, we considered bosonic canonical commutation relations.
\section{Quantum Fisher Information and why the measure matters}
\label{app:fisher:info}
Let us assume that the dynamics of a system described by the state $\hat{\rho}$ are Markovian, and that the state evolves according to
\begin{equation}
    \frac{d\hat{\rho}}{dt} = \mathcal{L}\left[\hat{\rho} \right]\,,
\end{equation}
where $\mathcal{L}$ is the generator of the dynamics and takes the Lindblad form.  
Following~\cite{Styliaris2020}, we define the symmetries of the Lindbladian as the set of superoperators that commute with $\mathcal{L}$
\begin{equation}
    \mathrm{Sym}(\mathcal{L}) := \left\{ \mathcal{M} \in \mathscr{B}(\mathscr{B}(\mathscr{H})) \;\middle\vert\; [\mathcal{M}, \mathcal{L}] = 0 \right\}\,,
\end{equation}
where $\mathscr{B}(\mathscr{B}(\mathscr{H}))$ denotes the set of bounded linear operators acting on the space of bounded linear operators on the finite-dimensional Hilbert space $\mathscr{H}$.

Here, we focus on the special case where the symmetry superoperator is the Lindbladian itself, i.e., $\mathcal{M} = \mathcal{L}$. In this case, any quantum Fisher information with respect to the time parameter serves as a monotone under the evolution generated by $\mathcal{L}$.
The family of quantum Fisher information (QFI) about the time parameter $t$ can be written in the form
\begin{equation}
    I_f(t) = \sum_{x, y} \frac{\abs{\braket{x(t)\abs{\,\mathcal{L} [\hat{\rho}_t]\,}y(t)}}^2}{p_x(t)\, f(p_y(t)/p_x(t))},
\end{equation}
where the density matrix is expressed in its spectral decomposition $\hat{\rho} = \sum_x p_x \ket{x}\!\bra{x}$. The set $\{p_x\}$ defines a discrete probability distribution over eigenstates. The function $f$ characterizes the particular QFI measure within the family and is required to satisfy three properties:  
1) operator monotonicity: for any positive semidefinite operators $A$ and $B$ such that $A \leq B$, it holds that $f(A) \leq f(B)$;  
2) self-inversiveness: $f(x) = x f(1/x)$ for all $x > 0$;  
3) normalization: $f(1) = 1$.

Common examples of such QFI measures include the symmetric logarithmic derivative (SLD) QFI with $f_{\mathrm{SLD}}(x) = \frac{x + 1}{2}$, which is minimal in this family, the Wigner-Yanase (WY) QFI with $f_{\mathrm{WY}}(x) = \frac{1}{4}(\sqrt{x} + 1)^2$, and the harmonic mean (HM) QFI with $f_{\mathrm{HM}}(x) = \frac{2x}{x + 1}$, which is the maximal element in this class.

For a more detailed discussion of these QFI families and their properties, see~\cite{Scandi2024}.

\begin{figure}
    \centering
\includegraphics[width=\linewidth]{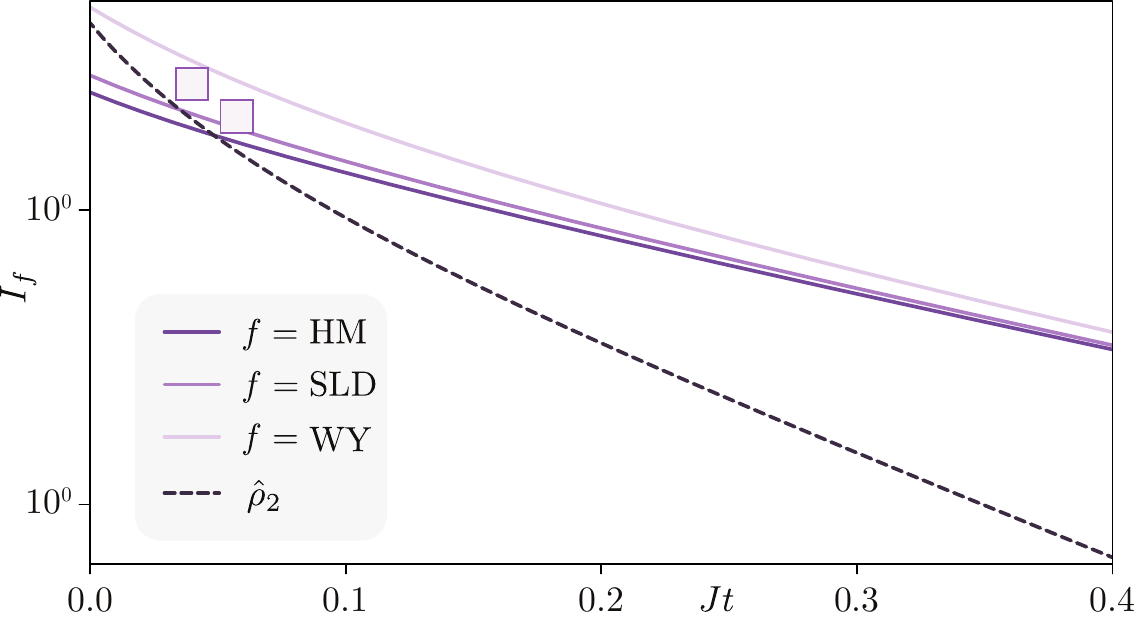}
    \caption{Quantum Fisher information $I_f$ for different monotones during the thermal relaxation of a single qubit. The initial states are unitarily connected, with one being incoherent (dashed line) and the other coherent (full lines) in the energy eigenbasis. Due to the structure of the Davies generator, the incoherent state's dynamics and corresponding $I_f$ are independent of the choice of $f$, while the coherent state's behavior depends strongly on the monotone. The SLD and WY monotones both exhibit an Mpemba crossing at distinct times, whereas the HM monotone does not display a crossing. Parameters: $(r_x, r_y, r_z) = (0.75, 0, 0.19)$, $J = 1$, $\beta = 0.1J^{-1}$, $\gamma = 1J$, $\omega = 5J$.}
\label{fig:qfi:1qubit}
\end{figure}

As discussed in~\cite{Styliaris2020}, this perspective can be naturally integrated into the framework of quantum resource theories, particularly those in which the free operations are generated by the dynamical map $\mathcal{E}_t := \exp(t\mathcal{L})$. In such a setting, the resource is identified with the degree of non-stationarity, as previously defined.

It is crucial to emphasize, however, that monotones associated with a given symmetry $\mathcal{M}$ do not, in general, induce a total order on quantum states. Specifically, if a monotone $M_a$ satisfies $M_a(\hat\rho_1) > M_a(\hat\rho_2)$ for two states $\hat\rho_1$ and $\hat\rho_2$, this does not necessarily imply that another monotone $M_b$ associated with the same symmetry will also yield $M_b(\hat\rho_1) > M_b(\hat\rho_2)$. In fact, we will present an explicit counterexample illustrating this point. Moreover, we explore the consequences of this monotone dependence in the context of crossing times, particularly as it pertains to (quantum) Mpemba behavior.

To this end, we revisit the thermal quantum Mpemba effect that emerges in the pure relaxation dynamics of a single qubit coupled to a thermal bath, as discussed in \cref{sec:example:davies:map:thermal} of the main text.  
We consider the relaxation behavior of two distinct, but unitarily connected, initial states: one of which is coherent in the energy eigenbasis, while the other ($\hat\rho_2$) is not. Owing to the structure of the Davies generator, the evolution of populations and coherences in the energy basis decouples. Furthermore, the quantity $I_f$ naturally splits into two contributions: one incoherent and one coherent, defined with respect to the instantaneous eigenbasis of the state.

Given these two facts, it is straightforward to see that for the incoherent state, only the incoherent component contributes to $I_f$. Importantly, this contribution is independent of the choice of the function $f$, as illustrated by the dashed grey line in \cref{fig:qfi:1qubit}.  
In contrast, the initially coherent state yields an $f$-dependent contribution, as shown by the colored lines in \cref{fig:qfi:1qubit}. Notably, the monotones $I_{\mathrm{SLD}}$ and $I_{\mathrm{WY}}$ exhibit a Mpemba crossing, though at distinct crossing times, while $I_{\mathrm{HM}}$ does not show any crossing. This demonstrates that both the presence of an Mpemba crossing and the associated crossing time are strongly dependent on the specific choice of monotone, even when the underlying resource is held fixed.
\section{From global G-invariance to local G-covariance}
\label{app:global:invariance:local:covariance}
In this section we demonstrate how a global symmetry of a joint system-environment unitary evolution induces covariance of the reduced system dynamics under the same symmetry group, if the environment is prepared in a symmetric state. 
Let us consider a $\mathrm{G}$ group, $\hat T\in \mathscr{B}(\mathscr{H}_s\otimes \mathscr{H}_e)$ a $\mathrm{G}$-invariant unitary, and $\hat \pi_e\in \mathscr{S}(\mathscr{H}_e)$ a $\mathrm{G}$-invariant state. 
Then, the map
\begin{equation}
    \mathcal{E}[\cdot] = \Tr_e[\hat T ([\cdot] \otimes \hat \pi_e) \hat T^\dagger]
\end{equation}
will be $\mathrm{G}$-covariant, namely
\begin{equation}
    \mathcal{E} [\hat U_{g,s}\, \hat \rho\, \hat U_{g,s}^\dagger ] = \hat U_{g,s}\, \mathcal{E}[\hat \rho] \, \hat U_{g,s}^\dagger \quad \forall\,\hat \rho\in \mathscr{S}(\mathscr{H}_s)\,,
\end{equation}
where $\hat U_{g,s}$ is the unitary representation of $\mathrm{G}$ on $\mathscr{H}_s$.\\
To prove this let us consider a unitary representation of the group $\hat{U}_g = \hat{U}_{g,s}\otimes\hat{U}_{g,e}$ on $\mathscr{H}_s\otimes \mathscr{H}_e$ and begin from the left-hand side
\begin{equation}
    \mathcal{E} [\hat U_{g,s}\, \hat \rho\, \hat U_{g,s}^\dagger] =\Tr_e[\hat T (\hat U_{g,s}\, \hat \rho\, \hat U_{g,s}^\dagger \otimes \hat \pi) \hat T^\dagger]\,.
\end{equation}
By $\mathrm{G}$-invariance we will have $[\hat T, \hat{U}_{g,s}\otimes\hat{U}_{g,e}]=0\, \forall\,g \in \mathrm{G}$, so
\begin{equation}
    \hat T (\hat U_{g,s}\, \hat \rho\, \hat U_{g,s}^\dagger \otimes \hat \pi) T^\dagger = \hat U_{g,s}\otimes \hat U_{g,e }\, (\hat \sigma)\, \hat U_{g,s}^\dagger\otimes \hat U_{g,e}^\dagger\,,
\end{equation}
where $\hat \sigma = \hat T (\hat \rho \otimes \hat \pi_e) \hat T^\dagger$, where we used $\hat \pi_e = \hat U_{g,e}\, \hat \pi_e\, \hat U_{g,e}^\dagger$.
Hence
\begin{equation}
    \mathcal{E} [\hat U_{g,s}\, \hat \rho\, \hat U_{g,s}^\dagger] =\Tr_e[\hat U_{g,s}\otimes \hat U_{g,e }\, (\hat \sigma)\, \hat U_{g,s}^\dagger\otimes \hat U_{g,e}^\dagger]\,.
\end{equation}
Since the partial trace over the environment commutes with unitaries on the environment, one gets
\begin{equation}
\begin{split}
    \Tr_e[\hat U_{g,s}\otimes \hat U_{g,e }\, (\hat \sigma)\, \hat U_{g,s}^\dagger\otimes \hat U_{g,e}^\dagger] &= \hat U_{g,s} \Tr_e[\hat \sigma]\hat U_{g,s}^\dagger \\
    &= \hat U_{g,s} \mathcal{E} [\hat \rho] \hat U_{g,s}^\dagger\,.
\end{split}
\end{equation}
Thus
\begin{equation}
    \mathcal{E} [\hat U_{g,s}\, \hat \rho\, \hat U_{g,s}^\dagger ] = \hat U_{g,s} \mathcal{E} [\hat \rho] U_{g,s}^\dagger\,,
\end{equation}
showing that $\mathcal{E}$ is $\mathrm{G}$-covariant whenever $\hat U$ is $\mathrm{G}$-invariant and $\hat \pi$ is $\mathrm{G}$-invariant.
%
%
%

\FloatBarrier
\bibliography{references}
\end{document}